\documentclass[journal=jacsat,manuscript=article]{achemso}

\usepackage{chemformula} 
\usepackage[T1]{fontenc} 
\usepackage{graphicx}
\usepackage{amsmath}
\usepackage{amssymb}
\usepackage{upgreek}
\usepackage{textcomp}
\usepackage{gensymb}
\usepackage{xcolor}
\usepackage{caption}
\usepackage{subcaption}
\usepackage{comment}
\usepackage{multirow}
\usepackage[normalem]{ulem}

\author{Pavel~A.~Dmitriev}
\affiliation[imre]{Institute of Materials Research and Engineering, A*STAR (Agency for Science, Technology and Research), 2 Fusionopolis Way, $\#$08-03 Innovis, 138634 Singapore}
\alsoaffiliation[ntu]{LUMINOUS! Centre of Excellence for Semiconductor Lighting and Displays, School of Electrical and Electronic Engineering, School of Physical and Mathematical Sciences, School of Materials Science and Engineering, Nanyang Technological University, Singapore 639798, Singapore}

\author{Emmanuel~Lassalle}
\affiliation[imre]{Institute of Materials Research and Engineering, A*STAR (Agency for Science, Technology and Research), 2 Fusionopolis Way, $\#$08-03 Innovis, 138634 Singapore}
\email{Emmanuel_Lassalle@imre.a-star.edu.sg}

\author{Lu~Ding}
\affiliation[imre]{Institute of Materials Research and Engineering, A*STAR (Agency for Science, Technology and Research), 2 Fusionopolis Way, $\#$08-03 Innovis, 138634 Singapore}

\author{Zhenying~Pan}
\affiliation[imre]{Institute of Materials Research and Engineering, A*STAR (Agency for Science, Technology and Research), 2 Fusionopolis Way, $\#$08-03 Innovis, 138634 Singapore}

\author{Darren~C.~J.~Neo}
\affiliation[imre]{Institute of Materials Research and Engineering, A*STAR (Agency for Science, Technology and Research), 2 Fusionopolis Way, $\#$08-03 Innovis, 138634 Singapore}

\author{Vytautas~Valuckas}
\affiliation[imre]{Institute of Materials Research and Engineering, A*STAR (Agency for Science, Technology and Research), 2 Fusionopolis Way, $\#$08-03 Innovis, 138634 Singapore}

\author{Ramón~Paniagua-Dominguez}
\affiliation[imre]{Institute of Materials Research and Engineering, A*STAR (Agency for Science, Technology and Research), 2 Fusionopolis Way, $\#$08-03 Innovis, 138634 Singapore}

\author{Joel~K.~W.~Yang}
\affiliation[sutd]{Singapore University of Technology and Design, 8 Somapah Road, 487372, Singapore}
\alsoaffiliation[imre]{Institute of Materials Research and Engineering, A*STAR (Agency for Science, Technology and Research), 2 Fusionopolis Way, $\#$08-03 Innovis, 138634 Singapore}

\author{Hilmi~Volkan~Demir} 
\affiliation[ntu]{LUMINOUS! Centre of Excellence for Semiconductor Lighting and Displays, School of Electrical and Electronic Engineering, School of Physical and Mathematical Sciences, School of Materials Science and Engineering, Nanyang Technological University, Singapore 639798, Singapore}
\alsoaffiliation[bilkent]{Department of Electrical and Electronics Engineering, Department of Physics, UNAM – Institute of Materials Science and Nanotechnology, Bilkent University, Ankara 06800, Turkey}

\author{Arseniy~I.~Kuznetsov}
\affiliation[imre]{Institute of Materials Research and Engineering, A*STAR (Agency for Science, Technology and Research), 2 Fusionopolis Way, $\#$08-03 Innovis, 138634 Singapore}
\email{Arseniy_Kuznetsov@imre.a-star.edu.sg}

\title{Hybrid Dielectric-Plasmonic Nanoantenna
with Multi-Resonances for Subwavelength Photon Sources}

\abbreviations{}

\begin{document}





\begin{abstract}
        The enhancement of the photoluminescence of quantum dots induced by an optical nanoantenna has been studied considerably, but there is still significant interest in optimizing and miniaturizing such structures, especially when accompanied with an experimental demonstration. 
        Most of the realizations use plasmonic platforms, and some also use all-dielectric nanoantennas, but hybrid dielectric-plasmonic (subwavelength) nanostructures have been very little explored.
		In this paper, we propose and demonstrate single subwavelength hybrid dielectric-plasmonic optical nanoantennas coupled to localised quantum dot emitters that constitute efficient and bright unidirectional photon sources under optical pumping. To achieve this, we devised a silicon nanoring sitting on a gold mirror with a $10\,$nm gap in-between, where an assembly of colloidal quantum dots is embedded. Such a structure supports both (radiative) antenna mode and (non-radiative) gap mode resonances, that we exploit for the dual purpose of out-coupling the light emitted by the quantum dots into the far-field with out-of-plane directivity, and for enhancing the excitation of the dots by the optical pump. Moreover, almost independent control of the resonance spectral positions can be achieved by simple tuning of geometrical parameters such as the ring inner and outer diameters, allowing to conveniently adjust these resonances with respect to the quantum dots emission and absorption wavelengths. Using the proposed architecture, we obtain experimentally average fluorescence enhancement factors up to $654\times$ folds mainly due to high radiation efficiencies, and associated to a directional emission of the photoluminescence into a cone of $\pm 17\degree$ in the direction normal to the sample plane.
        We believe the solution presented here to be viable and relevant for the next-generation of light-emitting devices.
		
\end{abstract}

\section{Keywords}
fluorescence enhancement, Purcell effect, optical nanoantenna, 
        nanoparticle-on-mirror, 
        nanopatch antenna, 
        quasi-normal modes

\section{Introduction}
	
    	With the constant effort for miniaturizing light-emitting devices, new nanoscale light-emitting diodes (also known as "nano-LEDs") and on-chip single photon sources with subwavelength dimensions, typically below $1\,\mu$m in the visible range, are more and more on-demand, for applications like advanced display, optical communications or quantum technologies. A promising way to create efficient and bright photon sources is to enhance the photoluminescence (fluorescence) of quantum emitters (such as quantum dots) with optical nanoantennas, which are optically resonant nanostructures designed to control and boost the light-matter interactions at the nanoscale through extreme confinement of the light field \cite{koenderink2017single}. This process is otherwise very inefficient due to impedance mismatch between the optical cross-sections of quantum emitters and the wavelength of light \cite{hugall2018plasmonic}.  
    	    	
    	In general, fluorescence enhancement can be achieved via three different mechanisms \cite{koenderink2017single}, such as local enhancement of the pump field intensity to enhance the excitation of the fluorescent emitters by the incident pump, modifying the electromagnetic environment via the local density of optical states (also known as ``Purcell effect'') to make the light emission process faster and/or more efficient, and shaping the emitted fluorescence towards the collection system to increase the collected signal. The optical nanoantennas used for fluorescence enhancement usually aim to leverage on combining part of --- if not all --- these mechanisms.

    	To date, the platforms that provide the highest fluorescence enhancements are plasmonic nanoparticle-on-mirror (also known as nanopatch optical antennas)\cite{baumberg2019extreme}, in which the quantum emitters are embedded in the narrow gap formed between a subwavelength plasmonic nanoparticle and a metallic mirror. This architecture was first pinpointed theoretically for the large Purcell factor they can provide \cite{esteban2010optical}, in order to dramatically accelerate the spontaneous emission process of the quantum emitters. It was later realized experimentally \cite{akselrod2014probing}, with many subsequent realizations \cite{hoang2015ultrafast,akselrod2015leveraging,akselrod2016efficient,hoang2016ultrafast, sugimoto2018hybridized}. However, these plasmonic nanopatch antennas work best for sub-$10\,$nm gap, which is suitable for the purpose of enhancing the fluorescent signal from small molecules, but not for e.g. modern colloidal quantum dots which are much more relevant for the creation of photon sources due to their stability, but are also much bigger with typical diameters $\sim 10\,$nm, and thus cannot be integrated in such small gaps \cite{hugall2018plasmonic}. Moreover, they suffer from absorption losses of their metal parts, that ideally must be reduced in order to improve the radiative efficiency of these nanoantennas (also called quantum efficiency or quantum yield), defined as the probability that the excitation of the quantum emitters results in a photon emitted in the far-field, which is estimated between $20-50\,\%$ \cite{koenderink2017single}. 
    	
    	As an alternative solution, all-dielectric nanoantennas, made of low-loss and high-refractive index materials, ensure higher radiative efficiencies and more flexibility for controlling the emission directivity and enhancing the emission of magnetic emitters\cite{bidault2019dielectric}. Nevertheless, despite providing in theory relatively high enhancement factors in comparison with plasmonic antennas for larger gaps \cite{albella2013low}, these realizations are usually far from the record-breaking performances of plasmonic nanopatch antennas \cite{regmi2016all, cambiasso2018surface}.  This field has also mainly been driven by single molecule microscopy rather than photonic integration.

    	%
    	%
    	Hybrid dielectric-plasmonic structures aim to take the best of both worlds, that is the field enhancement and confinement of plasmonic structures and the low losses and greater flexibility of dielectric structures to tune the resonances and shape the directivity\cite{lepeshov2019hybrid,barreda2021applications}. 
    	Most of them typically involve micron-scaled hybrid bullseye antennas\cite{livneh2015efficient,livneh2016highly,andersen2018hybrid,stella2019enhanced} to demonstrate high directivity, but such systems cannot be shrunk down to subwavelength sizes by nature. More recently, a particularly interesting design was put forward in the context of nanoparticle-on-mirror antennas with subwavelength size, consisting in a dielectric nanodisk sitting on top of a metallic mirror, with, in theory, high Purcell factor, quantum efficiency, and directional emission \cite{yang2017low}.	Two subsequent experimental works demonstrated the relevance and potential of such a hybrid platform by enhancing the photoluminescence of quantum dots located in the gap formed between a silicon nanosphere and a gold mirror \cite{sugimoto2018broadband, yang2019greatly}, essentially leveraging on the Purcell effect to boost the emission process, but without considering directivity control.
    	The absence of more experimental demonstrations with subwavelength nanoantennas could be because the integration of emitters with the nanoantenna is on its own a very complex problem, usually requiring either precise manipulation\cite{makarov2018nanoscale,bek2008fluorescence,schell2011scanning,schroder2011fiber,werschler2018efficient,kolchin2015high} or precise measurement of existing emitters location\cite{liu2021nanoscale}.
    
         In this work, we design a silicon hollow disk (i.e. a nanoring) antenna sitting on top of a gold mirror, forming a nanogap of about $\sim 10\,$nm, within which colloidal quantum dots are embedded. We thus obtained self-localised emitters, present underneath the nanoantenna only, alleviating the problem of quantum emitters localisation. Silicon nanorings present strong scattering properties, with geometrically tunable resonances (also called ``Mie resonances'') that provide control over the scattering strength and directivity, with potential ability to tailor the emission of electric or magnetic dipole emitters\cite{kuznetsov2020antenna,kuznetsov2019antenna,feng2016all,zenin2020engineering}, but it is generally difficult to  efficiently couple an assembly of emitters. Our hybrid nanopatch antenna allows us to achieve an efficient coupling due to the presence of two classes of resonant modes: strongly radiative modes that also provide out-of-plane directivity (called "antenna modes" hereafter), similar to the standalone nanoring resonances but with much better coupling strengths with the QDs due to the presence of the mirror; and weakly radiative modes confined in the gap between the particle and the mirror\cite{filter2012circular,yang2017low} (called "gap modes" hereafter; these two classes of modes are also discussed in related configurations, see e.g. \cite{esteban2015morphology}).
        We also find that these two types of modes can be spectrally tuned, almost independently from each other, by simply varying the nanoring geometrical parameters, such as its inner and outer diameters.
    	By carefully designing these parameters to match the antenna mode and gap mode resonances with the emission wavelength of our quantum dots and the excitation wavelength of the pump laser, respectively, we demonstrate experimentally a directional fluorescence enhancement over $650$ into a cone of $\pm 17\degree$ in the direction normal to the surface, compared to the rather isotropic emission of quantum dots on gold film. We clearly identify the main mechanisms responsible for the total enhancement, by the use of analytical and quasi-normal mode frameworks in conjunction with full-waves simulations, to be a combination of radiative, excitation, and directivity enhancements.
    	Overall, the hybrid dielectric-plasmonic nanoantenna reported here presents multi-resonances with strong local field confinement, scattering and directivity properties, that can be used with the association of an active medium (e.g. quantum dots) to create efficient, bright and directional photon sources.
    		
	\section{Design and fabrication}
	
\begin{figure*}[t]
    		\centering
    		\includegraphics[width=0.9\linewidth]{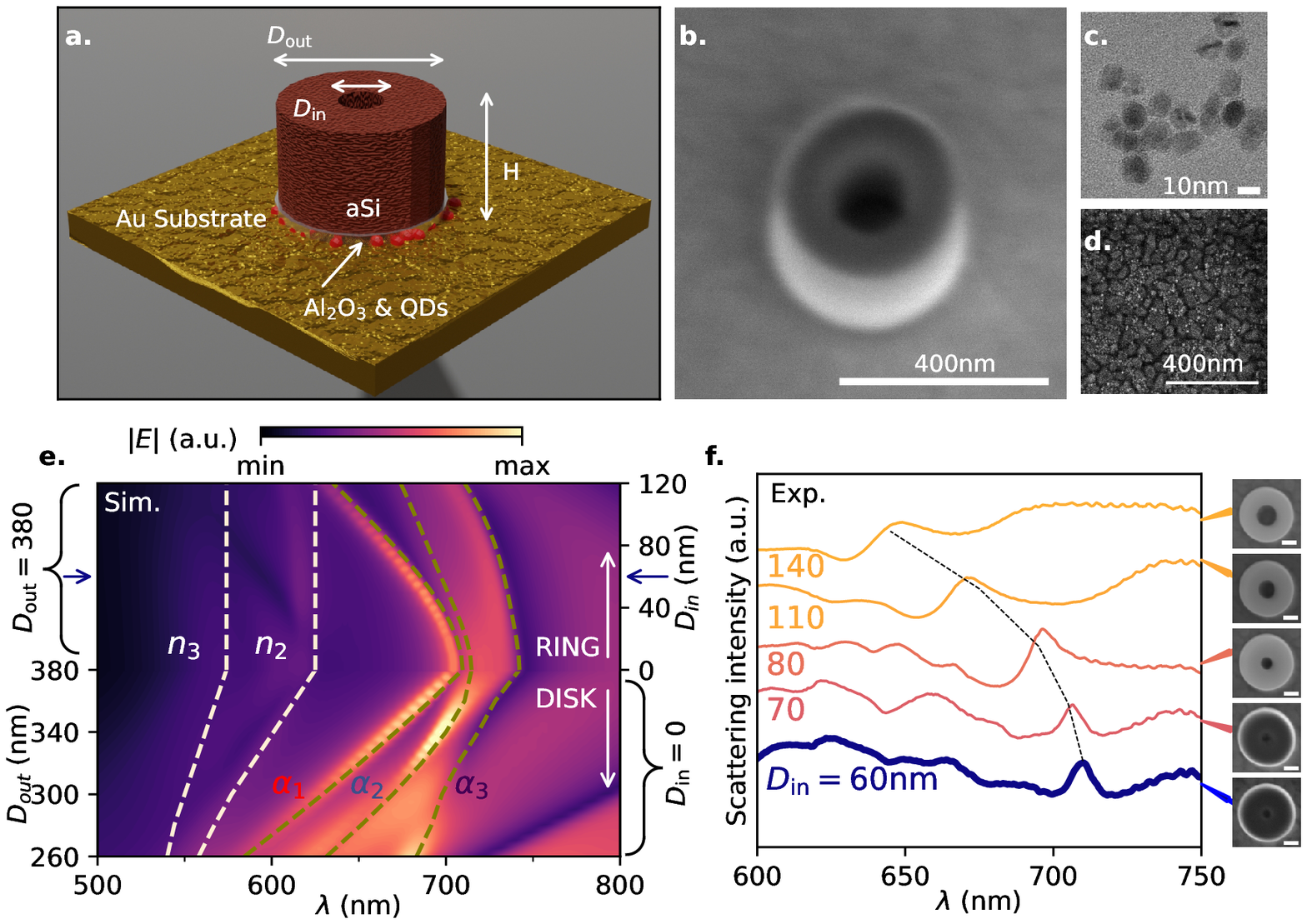}
    		\caption{\emph{Design and fabrication of hybrid dielectric-plasmonic nanoantennas.} \textbf{a.} Artist's impression of the aSi nanoring of height $H$, outer diameter $D_\text{out}$ and inner diameter $D_\text{in}$, placed on a Au substrate, with a Al$_2$O$_3$ spacer layer containing the embedded QDs (represented by the red dots). \textbf{b.} Tilted SEM image of a typical fabricated nanoantenna. The scale bar represents $400\,$nm. \textbf{c.} TEM image of the  CdSe/ZnS QDs. The scale bar represents $10\,$nm.
    		\textbf{d.} SEM image of the Au substrate coated by a thin layer of Al$_2$O$_3$ itself covered with spin-coated QDs. The scale bar represents $400\,$nm.
    		\textbf{e.} Simulated electric field in nanoantenna gap, for varying $D_\text{out}$ with fixed $D_\text{in} = 0\,$nm (disk case, lower half) or varying $D_\text{in}$ with fixed $D_\text{out}=380\,$nm (ring case, upper half). In all cases $H=230\,$nm. Olive green dashed lines show calculated resonance wavelengths of three antenna modes labelled $\alpha_1$, $\alpha_2$, $\alpha_3$. White dashed lines show calculated resonance wavelengths of second and third order gap plasmon modes labelled $n_2$, $n_3$. Blue arrows on the sides show the case $D_\text{in}=60\,$nm (chosen to be discussed in the rest of this work and called "Antenna"). \textbf{f.} Experimental scattering intensity spectra measured for different nanoantennas having approximately same $D_\text{out}=380\,$nm and $H=230\,$nm, but increasing $D_\text{in}$ (from bottom to top). The insets show the corresponding SEM images of the fabricated nanoantennas. The scale bars represent $100\,$nm. The scattering of the "Antenna" (case $D_\text{in}=60\,$nm) is highlighted by a thicker line. Dashed black line is a guide-to-the-eye showing the evolution of one of the resonances with changing inner hole diameter. 
            }
    		\label{fig1_designsalad}
	    \end{figure*}

    	The nanoantenna design, as shown schematically in Fig.~\ref{fig1_designsalad}a, consists of an amorphous silicon (aSi) nanoring sitting on a gold (Au) mirror substrate, with an alumina (Al$_2$O$_3$) spacer in between where the quantum emitters are embedded. Scanning electron microscopy (SEM) image of a fabricated sample is shown in Fig.~\ref{fig1_designsalad}b. We use CdSe/ZnS quantum dots (QDs) as quantum emitters, which have a strong electric dipole transition, with an emission centered around $650\,$nm\cite{lim2014influence}. 	A transmission electron microscope image of the QDs is shown in Fig.~\ref{fig1_designsalad}c, with QD sizes of the order of $\sim 10\,$nm. 
    	Their photoluminescence and absorption spectra measured in solution or on top of a glass substrate are given in Supp. Info. Section~1, Fig.~S1. 
    	
    	A basic model of their energy levels can be found in Supp. Info. Section~1, Fig.~S2, which constitutes the fluorescence model used to interpret our experimental results, and the basis on which most fluorescence experiments are analysed. In a reference configuration (chosen to be e.g. QDs deposited on a substrate), we associate a rate to each transition between different energy levels, namely an excitation rate $\gamma_\text{exc}^0$ induced by the excitation from a laser pump at the excitation wavelength $\lambda_\text{exc}$ (and directly proportional to its intensity), an intrinsic non-radiative decay rate $\gamma_\text{nr}^0$ due to non-radiative decay channels present in the emitter (such as vibrations or other non-radiative relaxation channels) and a radiative decay rate $\gamma_\text{r}^0$ associated to the spontaneous emission of a photon of emission wavelength $\lambda_\text{em}$ (Fig.~S2a). One can then characterize the emitter by its intrinsic quantum yield, defined as $\text{QY}^0 \equiv \gamma_\text{r}^0/(\gamma_\text{r}^0 + \gamma_\text{nr}^0)$, which quantifies the radiation efficiency (i.e. the probability that the excitation of the emitter actually results in the emission of a photon in the far field). 
        In the nanoantenna configuration, we assume that its presence does not modify the intrinsic non-radiative decay rate $\gamma_\text{nr}^0$, but does modify the other rates, which becomes $\gamma_\text{exc}^0\rightarrow\gamma_\text{exc}(\mathbf{r})$ for the excitation rate and $\gamma_\text{r}^0\rightarrow\gamma_\text{r}(\mathbf{r})$ for the radiative decay rate (Fig.~S2b). In addition, the nanoantenna introduces a new decay channel (highlighted by the dashed line in Fig.~S2b), which is the possibility that the emitted photon gets absorbed by the nanoantenna and lost in terms of heat, for which we associate an absorption decay rate $\gamma_\text{abs}(\mathbf{r})$. Note that the $\mathbf{r}$ dependence meaning that these rates now depend on the spatial location of the emitter --- while in the reference configuration they are usually not position-dependent. For a low-loss dielectric nanoantenna as the one used in this work, this absorption channel is expected to be small.
        The quantum yield of the emitter is thus modified as $\text{QY}(\mathbf{r}) = \gamma_\text{r}(\mathbf{r})/[\gamma_\text{r}(\mathbf{r}) + \gamma_\text{nr}^0 + \gamma_{\text{abs}}(\mathbf{r})]$.

    	The reference configuration considered in this work consists in a monolayer of QDs deposited on the gold mirror; we call it the ``Reference'' hereafter. An SEM image of the relatively homogeneous distribution of QDs after spin-coating can be seen in Fig.~\ref{fig1_designsalad}d.
        Note that in order to prevent quenching by the gold mirror and to protect from the subsequent CVD deposition of the aSi nanoantenna (see Methods section), the QDs were sandwiched between two ultra-thin layers of Al$_2$O$_3$ (we denote this configuration by Au/Al$_2$O$_3$/QDs/Al$_2$O$_3$). The overall thickness of the three layers is $\sim 10\,$nm.  
    	Importantly, the quantum yield of the QDs in the Reference situation was found to be $\text{QY}^0= 3.8\times 10^{-3}$, drastically decreased compared to the value for QDs in solution ($\text{QY}^0=0.29$). This observation is similar to what has been reported in Ref.~\cite{cihan2018silicon}, however in our case the decrease is not due to quenching by the gold mirror, but instead to some degradation that happens during the deposition of the second protective layer of Al$_2$O$_3$ (more details are given in Methods section). Finally, all the QDs which were not precisely located between the Au mirror and aSi nanoantenna were etched out; hence, we managed to create spatially self-aligned and localized QDs in a nanogap of $g\approx 10\,$nm, without requiring any complex emitter manipulation or characterization, similar to what was previously reported\cite{cihan2018silicon}.
    	    	
    	We first simulated the near-field optical response of our nanoantenna in Fig.~\ref{fig1_designsalad}e, which shows the norm of electric field $|\mathbf{E}|$ average in the middle plane of the nanogap (see Methods section for the simulation details), for a nanodisk with varying outer diameter $D_\text{out}$ (lower half), and for a nanoring with varying inner diameter (or hole diameter) $D_\text{in}$ (upper half). In both cases the height of the nanoantenna is fixed at $H=230\,$nm. One can see that while all modes are red-shifted as $D_\text{out}$ increases, the modes labelled by $\alpha_i$ are blue-shifted as $D_\text{in}$ increases, while the modes labelled by $n_i$ do not vary. As discussed in more details later, these modes pertained to two distinct classes: the modes $\alpha_i$ correspond to "antenna modes", which have strong radiative properties, and the modes $n_i$ correspond to "gap modes", mostly confined in the gap between the nanoantenna and the gold mirror, and much less radiative. As we will see in the following, because of the presence of these two classes of modes --- the antenna modes which are strongly sensitive to $D_\text{in}$, and the gap modes which are mostly sensitive to $D_\text{out}$ --- we are able to almost independently control the spectral positions of both types of resonances, allowing us to adapt the design to match the emission and excitation wavelengths of the QDs and the pump laser, respectively. Another route towards independent control these two types of resonances is to vary the height $H$ of the nanoantenna, as shown in Supp. Info. Section 2, Fig.~S3 (for a closely related configuration, also see \cite{esteban2015morphology}). In this work, we choose to set the height at $H=230\,$nm and the outer diameter $D_\text{out}=380\,$nm to have one main strongly radiative antenna mode around the QD emission wavelength of $650\,$nm (mode $\alpha_1$ in Fig.~\ref{fig1_designsalad}e) and a gap mode resonance that can be exploited for excitation enhancement around $570\,$nm (mode $n_3$ in Fig.~\ref{fig1_designsalad}e), while $D_\text{in}$ is used as "node" to spectrally tune the antenna resonances with respect to the gap mode resonances (as can be seen in the upper half of Fig.~\ref{fig1_designsalad}e).
	
    	We next designed and fabricated the aSi nanorings to have a fixed outer diameter of $D_\text{out}\approx 380\,$nm and a fixed height of $H\approx 230\,$nm, and varying inner diameters $D_\text{in}$ (ring hole) between $D_\text{in}\approx 60\,$nm to $D_\text{in}\approx 140\,$nm (SEM images of the nanorings are shown as insets in Fig.~\ref{fig1_designsalad}f). 
    	We characterized their far-field optical response by dark-field scattering measurements (see Methods section). Experimental scattering intensity spectra are shown in Fig.~\ref{fig1_designsalad}f and reveal that the scattering features are blue-shifted with increasing $D_\text{in}$, in agreement with the antenna modes behavior (while the gap modes are not really seen in the far-field as they do not radiate much).
    	
    	\section{Quasi-normal mode analysis of the antenna modes} 
    	 
    	In order to better understand the features appearing in the near-field spectra of Fig.~\ref{fig1_designsalad}e, we computed the resonant modes $\mathbf{E}_\alpha$ of the system as well as their complex eigenwavelengths $\lambda_\alpha=\lambda_\alpha'+\mathrm{i}\,\lambda_\alpha''$, with $\alpha$ labelling the mode, using quasi-normal mode (QNM) calculations \cite{https://doi.org/10.1002/lpor.201700113} (see Methods section for the calculation details). We identified that the main features appearing in the near-field spectra are associated to the excitation of three particular antenna modes, labelled $\alpha_1$, $\alpha_2$ and $\alpha_3$ (the real part of their eigenwavelengths $\lambda_\alpha'$, called resonance wavelength, is shown in Fig.~\ref{fig1_designsalad}e by olive green dashed lines).
    	
\begin{figure*}[!ht]
    		\centering
    		\includegraphics[width=0.9\textwidth]{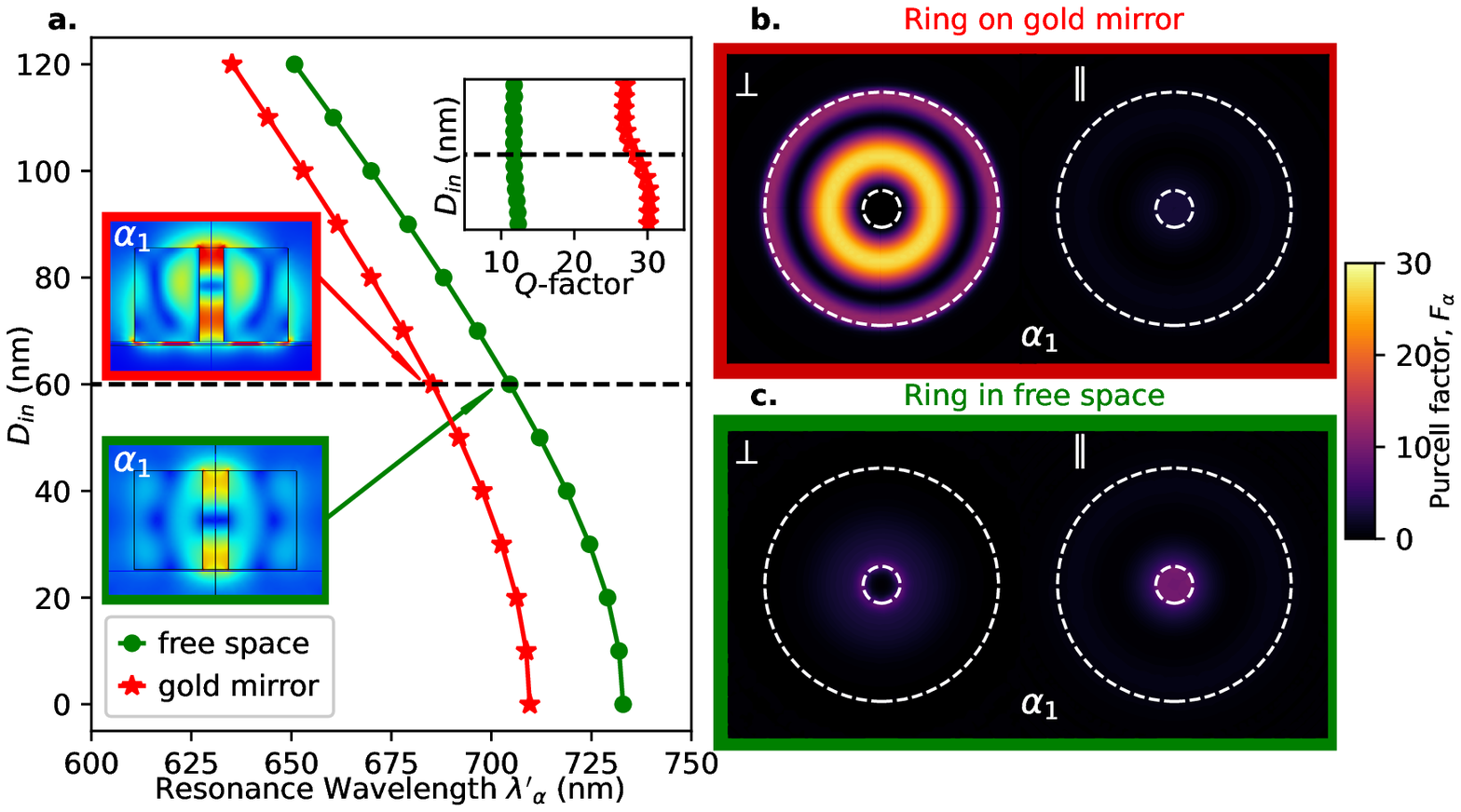}
    		\caption{\emph{Quasi-Normal Mode analysis of the nanopatch antenna mode $\alpha_1$ and comparison with an identical nanoring in free space.} \textbf{a.} Inner (hole) diameter $D_\text{in}$ \textit{vs.} resonance wavelength $\lambda_\alpha'$ of the QNM $\alpha_1$ for the nanoring on top of gold mirror (red dots, same as in Fig.~\ref{fig1_designsalad}e) and identical nanoring standing in free space (green dots). The top right inset shows the inner diameter $D_\text{in}$ \textit{vs.} $Q$-factor of the mode $\alpha_1$ in the two cases (same color code). The case of $D_\text{in}=60$ (chosen to be discussed in the rest of this work and called "Antenna") is also highlighted (horizontal black dashed lines). The two insets in the left side represent the norm of the QNM field $\alpha_1$ $|\mathbf{E}_\alpha|$ (in the vertical plane passing through the middle of the nanoantenna) for the nanoring with $D_\text{in}=60\,$nm on gold mirror (upper inset) and in free space (lower inset). Redder (bluer) colors represent high (low) values. \textbf{b.} Spatial distribution of the modal Purcell factor associated to the QNM $\alpha_1$, computed according to Eq.~(\ref{eq:PF}) in the horizontal plane located $5\,$nm below the nanoring, in the case of the nanoring on gold mirror with $D_\text{in}=60\,$nm (Antenna case). We discriminate between the out-of-plane oriented dipoles (symbol $\perp$) and the in-plane oriented dipoles (symbol $\parallel$, where in this case the Purcell factor is averaged over two orthogonal orientations). \textbf{c.} Same for nanoring in free space. The dotted white lines in \textbf{b.} and \textbf{c.} represent the projection of the outer and inner diameters of the nanoring in the horizontal plane.}
    		\label{fig2_modes}
    	\end{figure*}

    	For the nanoring, the mode $\alpha_1$ presents a higher $Q$-factor of about $Q\sim 30$ compared to the modes $\alpha_2$ and $\alpha_3$ with $Q\sim 10$ and $Q\sim 20$, respectively (calculated as $Q=\lambda_\alpha'/(2\lambda_\alpha'')$). Moreover, its spectral position is more shifted as with $D_\text{in}$ varies than for the two other QNMs, as can be seen in Fig.~\ref{fig1_designsalad}e, allowing for a higher degree of tunability. For these reasons, we choose the antenna mode resonance $\alpha_1$ for the purpose of enhancing the emission of the QDs. 
    	The field profile of this mode $|\mathbf{E}_\alpha|$, shown as insets in Fig.~\ref{fig2_modes}a, reveals that most of the electric field is located inside the hole of the nanoring, which explains why this mode is highly sensitive to the parameter $D_\text{in}$, which is not the case for the other modes $\alpha_2$ and $\alpha_3$ (not shown here). 
    	
    	It is interesting to make a comparison with the case of a standalone nanoring in free space; Fig.~\ref{fig2_modes}a shows the resonance wavelength vs $D_\text{in}$ in the case of the nanoring in free space (green dots).One can see that
    	that the presence of the metallic mirror (red dots, same as $\alpha_1$ in Fig.~\ref{fig1_designsalad}e), in addition to contributing to slightly blue-shifting its spectral position compared to the standalone case (green dots), contributes to more than doubling the $Q$-factor of the mode $\alpha_1$ (see inset in Fig.~\ref{fig2_modes}a with same color code). We also calculated the Purcell factor associated to this mode (called "modal Purcell factor"), for the nanoring in free space and in the presence of the metallic mirror, according to the formula:
    	 \begin{equation}
            F_\alpha(\mathbf{r})=\frac{6\pi c^3}{\omega'^3_\alpha}Q_\alpha\text{Re}\left(\frac{1}{V_\alpha(\mathbf{r})}\right)
            \label{eq:PF}
        \end{equation}
        with $c$ being the speed of light in vacuum. The modal Purcell factor quantifies the coupling of an emitter located at position $\mathbf{r}$ with QNM $\alpha$, and corresponds to the decay rate enhancement due to this mode compared to a homogeneous background (free space is considered here), for a perfect matching of the emission frequency of the emitter with the (real part of the) QNM eigenfrequency. The mode volume $V_\alpha(\mathbf{r})$ associated with this QNM, and appearing in Eq.~(\ref{eq:PF}), is given by the relation \cite{PhysRevLett.110.237401,https://doi.org/10.1002/lpor.201700113}:
        \begin{equation}
            V_\alpha(\mathbf{r})=\frac{1}{2\epsilon_0 \left(\mathbf{E}_\alpha(\mathbf{r})\cdot \mathbf{u}\right)^2}
            \label{eq:V}
        \end{equation}
        with $\epsilon_0$ being the vacuum permittivity. One can see from Eq.~\ref{eq:V} that the volume quantifies the interaction between electric dipole emitters with dipole orientation along unit vector $\mathbf{u}$ and the QNM field $\mathbf{E}_\alpha(\mathbf{r})$ at the position of the emitter $\mathbf{r}=(x,y,z)$. The smaller the mode volume, the stronger the interaction\cite{PhysRevLett.110.237401,https://doi.org/10.1002/lpor.201700113}.
        
        Figs.~\ref{fig2_modes}b and c show the spatial distribution of the modal Purcell factor associated with the antenna mode $\alpha_1$, across the horizontal plane located $5\,$nm underneath the ring, in the case of a nanoring with $D_\text{in}=60\,$nm (which is the case discussed hereafter in the main text), for dipoles oriented out-of-plane (i.e. perpendicular to the plane, denoted by the symbol $\perp$) and dipoles oriented in-plane (i.e. parallel to the plane, denoted by the symbol $\parallel$). 
    	
        One can see that in the presence of the mirror (Fig.~\ref{fig2_modes}b), it is the \emph{out-of-plane} dipoles that mostly couple to this mode, within a much larger area (that forms a circular ``band'' surrounding the hole of the nanoring as seen in the left panel of Fig.~\ref{fig2_modes}b), with a maximum modal Purcell factor reaching a value of $F_\alpha=27.4$.
        For the standalone nanoring case (Fig.~\ref{fig2_modes}c), it is mostly the \emph{in-plane} dipoles that couple to this mode, and only within a small area (located right underneath the hole of the nanoring as seen in the right panel of Fig.~\ref{fig2_modes}c). In practice, since it is very challenging to control the position and orientation of the emitters precisely, the PL signal coming from the assembly of emitters is averaged out over emitters spatially distributed over the nanoantenna area with random dipole orientations. Therefore, QDs coupled to a nanoring on top of a metallic mirror is expected to give a significantly higher PL enhancement compared to the case of a standalone nanoantenna, due to higher coupling strengths and the more spatially ``extended'' coupling.

	   \section{Theoretical analysis of the gap modes}
	   
	It is not very convenient to identify the gap modes using QNM computations because in the spectral region around and below $600\,$nm, the number of QNMs is very large and they are not  spectrally well separated, which makes their analysis quite tedious. Instead, we use a more intuitive analytical approach to find the gap mode resonances, based on geometrical considerations only.
    As mentionned in \citet{yang2017low}, the gap resonances of our system can be understood as ``the surface plasmon of a planar multilayer metal-dielectric system restricted to specific quantized wavectors''. The cylindrical symmetry of our system implies that resonances can be labeled with indices $(n, m)$, enumerating field variations in the radial and azimuthal directions, respectively. The wavevectors $k_{mn}$ are ``quantized'' due to the geometry of the nanoantenna, which reflects the surface plasmon at its boundary, similarly to the modes of a Fabry-Perot type resonator\cite{nielsen2011continuous,filter2012circular,ciraci2013quasi,tserkezis2015hybridization}.
    
	The resonance wavelengths, denoted $\lambda_{mn}$, thus have a geometric origin and read (in the case of a disk)\cite{filter2012circular}:
	   \begin{equation}
	   \lambda_{mn}=\frac{2\pi}{k_{mn}}\quad\text{with}\quad k_{mn}D_\text{out}+\phi=2J_{mn}
	   \label{eq:FP}
	   \end{equation}
	   where $k_{mn}$ is the real part of the surface plasmon polariton wavevector $k$ (whose multilayer dispersion relation is given in Supp. Info. Section S3), $D_\text{out}$ is the diameter of the disk, $J_{mn}$ is the $n^\text{th}$ zero of the Bessel's function of the first kind of order $m$ with $m$ the azimuthal number, and $\phi$ is a reflection phase that depends upon the structural and material parameters\cite{nielsen2011continuous}.
	  These gap mode resonances are thus spectrally tunable by simply varying the diameter of the particle $D_\text{out}$. Moreover, since these gap modes are strongly confined inside the nanogap, they can be exploited for local enhancement of the pump field intensity, and hence for excitation enhancement of the QDs.

	 We make use of Eq.~(\ref{eq:FP}) in the case of the ring as a first approximation --- even if it strictly holds for a plain disk only --- to calculate the gap resonance wavelengths in our system. Due to the symmetry of the pump field that we will use in this work (normal incidence and linearly polarized), only the modes with $m=1$ can be excited. In Fig.~1e, we show the resonance wavelengths $\lambda_{12}$ and $\lambda_{13}$ (dashed lines in white color and labelled $n_2$ and $n_3$, respectively --- the mode $n=1$ being outside of the wavelength range of interest), obtained by applying Eq.~(\ref{eq:FP}) with $\phi=-\pi/2$ and $\phi=0$, respectively (reflection phase values were chosen to closely match these analytically calculated resonance wavelengths with the features appearing in the near-field spectra simulations of Fig.~1e; also, since these reflection phases are related to the extension of the plasmon field beyond the particle terminations\cite{nielsen2011continuous}, it should not be too surprising that different modes pick up different reflection phases). One can see indeed that there is almost no changes when increasing the inner diameter of the ring $D_\text{in}$, which justify the use of this equation even in the case of a ring (especially as inner diameter remains small).
	 In particular, for $D_\text{out}= 380\,$nm, we have the following resonance wavelengths: $\lambda_{12}=625\,$nm and $\lambda_{13}=574\,$nm.
	  In the following, we use $n_3$ mode to enhance the excitation of the QDs, because the spectral position of $n_2$ overlaps with the emission of the QDs (see Supp. Info. Fig.~S1), making it difficult to distinguish the emission from the pump in the PL signal.

	 \section{Fluorescence enhancement results} 
	 
	  In the following, we present the results obtained for the selected nanoring with $D_\text{in}=60\,$nm (shown by blue arrows line in Fig.~\ref{fig1_designsalad}e), which we call ``Antenna'' hereafter, for which we obtained the highest total fluorescence enhancement experimentally. To characterize the total fluorescence enhancement of the QDs in the Antenna situation compared to the Reference one (denoted by the superscript ``$^0$'' hereafter), we use the well established fact that, in the low excitation regime, the fluorescence enhancement per emitter (located at position $\mathbf{r}$) is proportional to the gains in excitation rate, collection efficiency and quantum yield \cite{aouani2011bright, akselrod2014probing,koenderink2017single, mertens2007plasmon}:
        \begin{equation}
            \text{EF}_\text{th}(\mathbf{r}) = \frac{\gamma_\text{exc}(\mathbf{r})}{\gamma_\text{exc}^0}\frac{\text{D}_\text{em}(\mathbf{r})}{\text{D}_\text{em}^0}\frac{\text{QY}(\mathbf{r})}{\text{QY}^0}
            \label{eq:ef}
        \end{equation}
        The excitation rate $\gamma_\text{exc}$ (previously introduced) is directly proportional to the local enhancement of the pump intensity at the position of the emitter $\mathbf{r}$, and depends therefore on the wavelength used for the pump laser $\lambda_\text{exc}$. The directivity (or collection efficiency) $\text{D}_\text{em}$, which corresponds to the collected signal into a given numerical aperture (NA), thus depends on the collection NA, denoted by  $\text{NA}_\text{col}$, and also on the wavelength of fluorescent emission $\lambda_\text{em}$. Finally, the emitter quantum yield $\text{QY}$ (previously introduced) also depends on the emission wavelength $\lambda_\text{em}$. 
        Note that all of these quantities are defined for a single electric dipole emitter at a given position $\mathbf{r}$ and also having a fixed orientation of its dipole moment along the unit vector $\mathbf{u}$, but for the sake of readability, we omit the dependence on the parameter $\mathbf{u}$ in the above quantities.
        Also, it is interesting to note from Eq.~(\ref{eq:ef}) that the total fluorescence enhancement $\text{EF}_\text{th}$ depends on the type of emitter used in the Reference situation through its intrinsic quantum yield $\text{QY}^0$, and is thus not an absolute figure-of-merit to characterize the performance of a given antenna.
        
        However, in our case where the QDs have a very small intrinsic quantum yield $\text{QY}^0\ll 1$, we have $\gamma_\text{r}^0\ll \gamma_\text{nr}^0$ (here $\gamma_\text{r}^0\sim 10^{-3}\,\gamma_\text{nr}^0$), and further assuming that the that the decay rate enhancement in the presence of the nanoantenna is such that $\gamma_\text{r}(\mathbf{r}) + \gamma_{\text{abs}}(\mathbf{r})\ll \gamma_\text{nr}^0$, we have $\tau(\mathbf{r})/\tau^0\approx 1$, where we further introduce the fluorescence lifetimes defined as $\tau(\mathbf{r})\equiv 1/[\gamma_\text{r}(\mathbf{r}) + \gamma_\text{nr}^0 + \gamma_{\text{abs}}(\mathbf{r})]$ and $\tau^0\equiv 1/(\gamma_\text{r}^0 + \gamma_\text{nr}^0)$ in the nanoantenna and reference situations, respectively. Using the fact that $\text{QY}(\mathbf{r})/\text{QY}^0=[\gamma_\text{r}(\mathbf{r})/\gamma_\text{r}^0][\tau(\mathbf{r})/\tau^0]$, we thus have $\text{QY}(\mathbf{r})/\text{QY}^0\approx \gamma_\text{r}(\mathbf{r})/\gamma_\text{r}^0$, and Eq.~(\ref{eq:ef}) can be recast into the following form, which is independent of the type of emitter and will be more convenient for our purpose of characterising the underlying mechanisms at play in our nanoantenna\cite{aouani2011bright}:
        
        \begin{equation}
            \text{EF}_\text{th}(\mathbf{r})        \approx\eta_\text{exc}(\mathbf{r},\lambda_\text{exc})\times\eta_\text{em}(\mathbf{r},\lambda_\text{em}, \text{NA}_\text{col})
            \label{eq:ef_bis}
        \end{equation}
        
        In Eq.~(\ref{eq:ef_bis}), we defined the quantities $\eta_\text{exc}\equiv\gamma_\text{exc}/\gamma_\text{exc}^0$ and $\eta_\text{em}\equiv(\text{D}_\text{em}/\text{D}_\text{em}^0)(\gamma_\text{r}/\gamma_\text{r}^0)$, which quantify the enhancements in excitation and emission, respectively, and we made the dependence in terms of $\lambda_\text{exc}$, $\lambda_\text{em}$ and $\text{NA}_\text{col}$ of the terms $\eta_\text{exc}$ and $\eta_\text{em}$ explicit.  
        Applying Eq.~(\ref{eq:ef_bis}) to the case of our Antenna to each QD and averaging over QDs positions and orientations gives a theoretical fluorescence enhancement factor of $\left<\text{EF}_\text{th}\right>=1263$, for an excitation source coming at normal incidence and linearly polarized, and for a collection in a single direction, i.e. the upward direction (see Methods section and Supp. Info. Section~S4 for more details). The excitation and emission wavelengths used in the simulations, that is $\lambda_\text{exc}=583\,$nm and $\lambda_\text{em}=680\,$nm, were obtained by optimizing the excitation and emission separately (see next sections).
        
\begin{figure*}[!ht]
        	\centering
        	\includegraphics[width=1.0\textwidth]{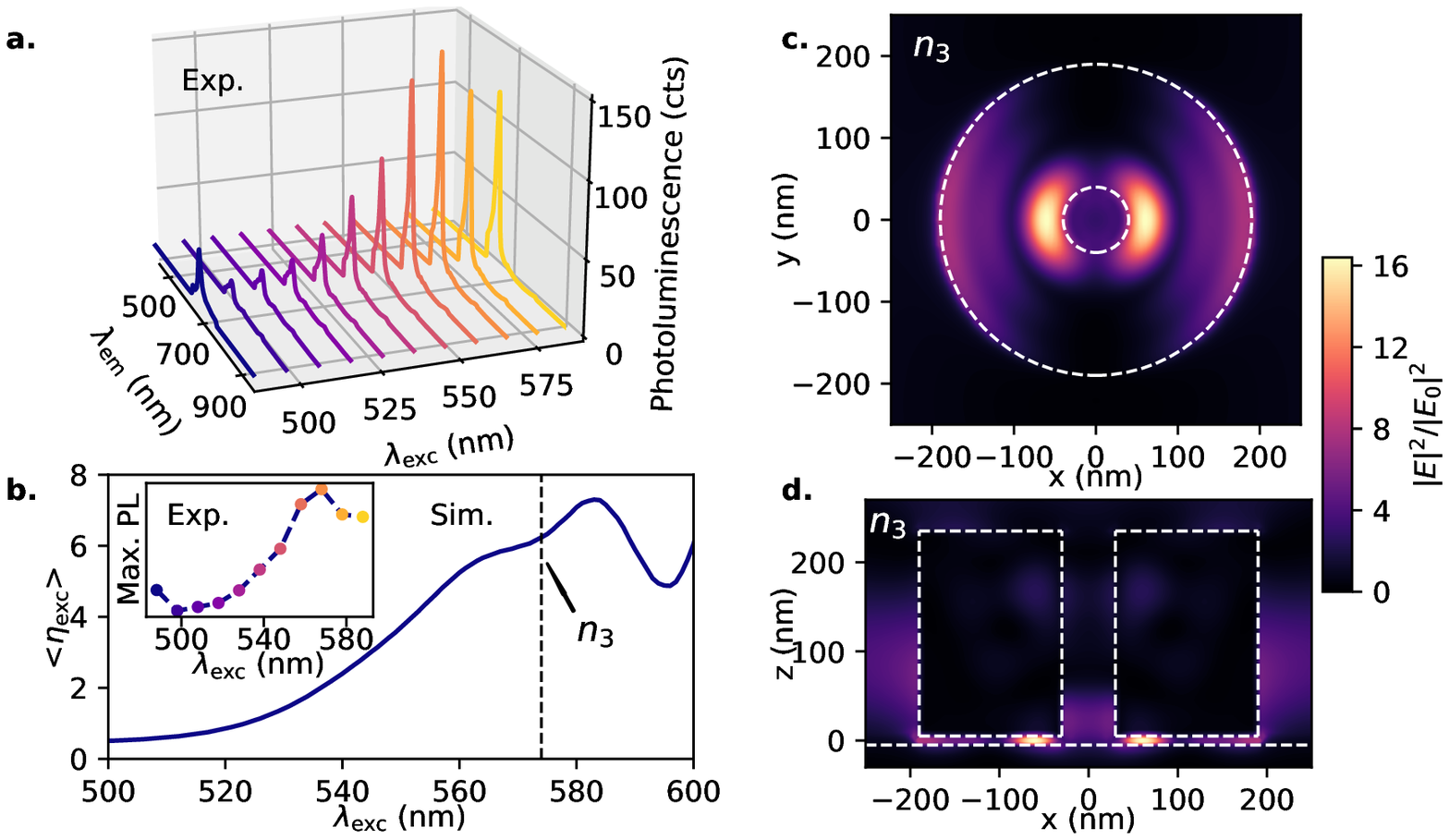}
        	\caption{\emph{Excitation enhancement.} \textbf{a.} Experimental spectra of PL counts (cts) in Antenna case, showing their dependence on the pump excitation wavelength $\lambda_\text{exc}$. The maximum emission peak is obtained for $\lambda_\text{exc}=570\,$nm. \textbf{b.} Simulated average excitation enhancement $\left<\eta_{\text{exc}}\right>_\text{th}$ for Antenna case as a function of the excitation wavelength $\lambda_\text{exc}$, computed according to Eq.~(\ref{eq:excitation_enhancement}) (see Methods section). The vertical dashed line denotes the theoretical resonance wavelength of the gap mode at $\lambda_\text{exc}=574\,$nm (third harmonic).
        	The inset shows the maxima of the PL spectra of \textbf{a.} (same color code) as a function of $\lambda_\text{exc}$, for better comparison with the simulation result (the dashed line connecting the points is guide-to-the-eye). \textbf{c.} and \textbf{d.} Simulated pump field intensity distribution in the horizontal cross-section passing through the middle of the nanogap (\textbf{c.}) and in the vertical cross-section passing through the middle of the nanoring (\textbf{d.}), respectively, at the maximum of excitation enhancement. In the simulations, the excitation source has its electric field linearly polarized along the $x$-direction, and comes at normal incidence, like in the experiment.}
        	\label{fig3_gapfield}
        \end{figure*}

        \emph{Excitation enhancement and gap resonance.} Experimentally, we first searched for the excitation wavelength $\lambda_\text{exc}$ that maximizes the PL. For that, we recorded the PL signal as we varied the pump wavelength $\lambda_\text{exc}$ from $488\,$ to $588\,$nm, while maintaining a constant pump power, by collecting the light radiated into air using an objective lens with $\text{NA} = 0.9$. Even though the QDs absorb shorter wavelength light more efficiently (as shown in Supp. Info. Section~S1), optimal pumping conditions for PL enhancement in the Antenna case were found to be at approximately $\lambda_\text{exc}=570\,$nm, as shown in Fig.~\ref{fig3_gapfield}a. In all subsequent measurements, we therefore fix the pump wavelength at $\lambda_\text{exc}=570\,$nm. 
    
       We confirmed with full-wave simulations that there is a maximum for the local field intensity around $570\,$nm (exactly at $583\,$nm), as shown in Fig.~\ref{fig3_gapfield}b (see Methods section for details about the full-wave simulations), which gives an excitation enhancement factor $\left<\eta_\text{exc}\right>_\text{th}=7.3$ --- the bracket denoting position and orientation averaging of the QDs. The spectral position of the maximum is in fair agreement with the third order gap mode resonance labelled $n_3$ predicted by the theory at $574\,$nm (vertical dashed line in Fig.~\ref{fig3_gapfield}b).
       Moreover, the field intensity in the horizontal cross-section located in the middle of the gap, and in vertical cross-section passing through the middle of the nanoring, shown in Figs~\ref{fig3_gapfield}c, d, respectively, reveal that intensity ``hot spots'' are formed within the nanogap. To better appreciate the match with the theory, we show in the Supp. Info. Section S3,  Fig.~S5 the vertical field component $E_z$, which matches the field profile of the gap mode expected in theory with symmetry $(n=3,m=1)$ (see Supp. Info. Section S3,  Fig.~S4a). This corroborates the fact that it is a gap mode resonance that is responsible for the excitation enhancement.
        Finally, note that the intensity distribution of this gap mode presents a rather good spatial overlap with the area within which the mostly coupled emitters to the antenna mode (exploited for \emph{emission} enhancement) shown in Fig.~\ref{fig2_modes}b.

\begin{figure*}[!ht]
        	\centering
        	\includegraphics[width=0.9\textwidth]{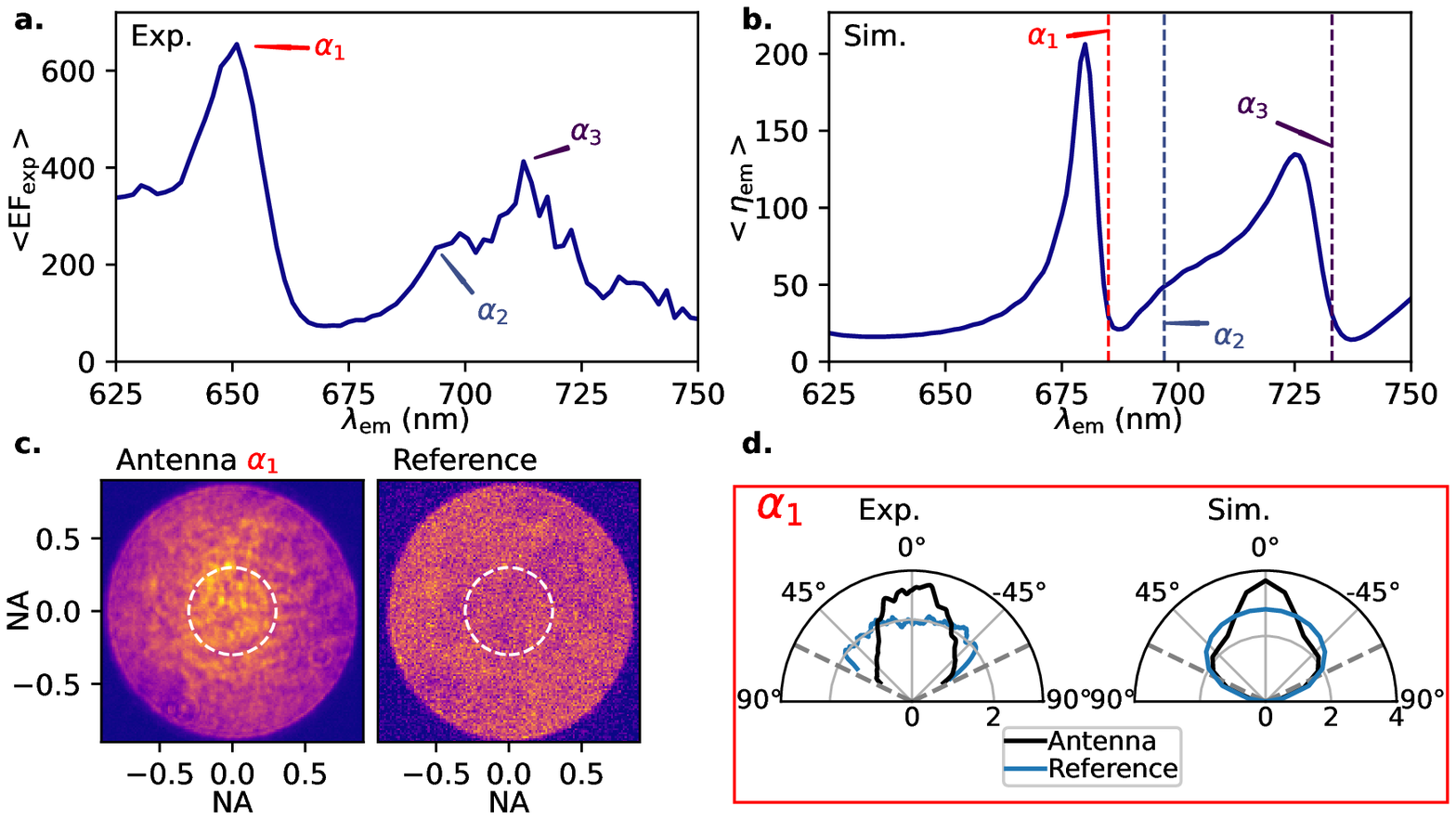}
        	\caption{
        	\emph{Emission enhancement.} 
        	\textbf{a.} Experimental average enhancement factor $\left<\text{EF}_\text{exp}\right>$ spectrum obtained after deconvoluting the Antenna PL from the Reference PL, according to Eq.~(\ref{eq:exp}), and restricting the PL integration over $\text{NA}_\text{col}=0.3$ (to quantify the directional emission enhancement). 
        	\textbf{b.} Simulated emission enhancement $\left<\eta_\text{em}\right>_\text{th}$ spectrum for Antenna computed in the upward direction. The vertical dashed lines denotes the resonance wavelengths of QNMs $\alpha_1$, $\alpha_2$ and $\alpha_3$ computed in Fig.~\ref{fig1_designsalad}f, enabling us to identify the main peaks with these resonances.
        	\textbf{c.} Experimental back focal plane (BFP) images of the emission PL of Antenna (left) and Reference (right) cases taken with a bandpass filter centered around $\lambda_\text{em}=650\,$nm (with a $30\,$nm bandwidth). The dotted circles represent $\text{NA}=0.3$. The intensities are not normalized.
        	\textbf{d.} Left: Experimental angular radiation patterns obtained as a linear cut of the BFP images in the Antenna (dark curve) and Reference cases (light blue curve) in \textbf{c.}. The grey dashed lines show the maximum collection angle corresponding to $\theta=\pm 64.2\degree$.
        	Right: Simulated angular radiation patterns in the Antenna (dark curve) and Reference cases (light blue curve), computed at the wavelength corresponding to the maximum peak ($\alpha_1$) in \textbf{b.}. The intensities are normalized such that the integration over all angles gives a value of $2\pi$.
        	}
        	\label{fig4_enhance}
        \end{figure*}

        \emph{Emission enhancement and antenna resonance.} 
        We quantify hereafter the emission enhancement in the upward direction, since our nanoantennas radiate mostly in the upward (out-of-plane) direction (see directivity patterns which are given later). For that, we choose to integrate the experimental angle-resolved PL (raw angle-resolved PL spectra can be found in Supp. Info. Section~5 Fig.~S7), obtained using back focal plane imaging technique (see Methods section), over a collection $\text{NA}_\text{col}=0.3$, which corresponds approximately to $\pm 17\degree$. Note that we make the choice to integrate over this collection NA (instead of considering strictly a single direction, i.e. the upward direction, like in the simulations) in order to average the noise present in the collection channel. Moreover, in order to obtain the experimental total fluorescence enhancement from these PL spectra, we deconvoluted the PL signal of the nanoantenna with the Reference PL, according to the formula:
        \begin{equation}
            \left<\text{EF}_\text{exp}\right>=\frac{I}{I_0}\frac{\mathcal{A}_0}{\mathcal{A}}
            \label{eq:exp}
        \end{equation}
        where $I$ (resp. $I_0$) is the PL intensity collected in the nanoantenna situation (resp. in the Reference situation) and $\mathcal{A}$ (resp. $\mathcal{A}_0$) is the area corresponding to the ring horizontal cross-section that reads $\mathcal{A}=\pi (D_\text{out}/2)^2$ with $D_\text{out}=380\,$nm where the QDs are located (resp. the area of the excitation spot $\mathcal{A}_0=\pi (D_\text{spot}/2)^2$ which is estimated to $D_\text{spot}\simeq 1.37\,\mu$m; see Supp. Info. Section~S6, Fig.~S8). 
        The obtained experimental total fluorescence enhancement spectrum is plotted in Fig.~\ref{fig4_enhance}a. One can see one main peak around the central emission wavelength of the QDs which is around $650\,$nm, at which we get a maximum total fluorescence enhancement factor of $\left<\text{EF}_\text{exp}\right>=654$. One can also see a secondary peak around $720\,$nm. 

        We corroborated this experimental total fluorescence enhancement spectrum with full-wave simulations of the emission enhancement $\left<\eta_\text{em}\right>$ in the upward direction (using reciprocity, see Methods section), because it corresponds to the direction of maximum directivity, and where the bracket denotes once again position and orientation averaging of the QDs. The simulated spectrum is shown in Fig.~\ref{fig4_enhance}b. One can see a fair agreement with the experiment, with also the presence of two main peaks, at $680\,$nm and $725\,$nm in the simulations. The shift of the resonances in the simulations compared to the experiment can be attributed to slight size variations in the height and outer diameter of the ring.
        Moreover, from the previous mode analysis shown in Fig.~\ref{fig1_designsalad}e, we identify these two resonances with the antenna modes $\alpha_1$ and $\alpha_3$, respectively, as highlighted in Fig.~\ref{fig4_enhance}b by vertical dashed lines. One can even see a weaker third resonance in-between the main two peaks, which we identify with the antenna mode $\alpha_2$ from Fig.~\ref{fig1_designsalad}e, as highlighted in Fig.~\ref{fig4_enhance}b. 
        The maximum emission enhancement at the main peak is $\left<\eta_\text{em}\right>_\text{th}=206.2$. Note that the value of the averaged total enhancement $\left<\text{EF}_\text{th}\right>=1263$ is not strictly equal to the average excitation enhancement $\left<\eta_\text{exc}\right>_\text{th}=7.3$ times the average emission enhancement $\left<\eta_\text{em}\right>_\text{th}=206.2$, which is totally normal and is due to the averaging process. Indeed, $\left<\text{EF}_\text{th}\right>\neq\left<\eta_\text{exc}\right>_\text{th}\times \left<\eta_\text{em}\right>_\text{th}$ in general for spatially inhomogeneous couplings.
        
        We show in Fig.~\ref{fig4_enhance}c (left panel) the experimental back focal plane image of the PL intensity at the wavelength $650\,$nm, coming from the main peak and associated to the QNM $\alpha_1$ (see Methods section for experimental details). One can see some directivity, compared to the rather isotropic emission from the Reference (right panel). It becomes even clearer when plotting the experimental angular radiation patterns --- obtained as a horizontal cut of Fig.~\ref{fig4_enhance}c --- in Fig.~\ref{fig4_enhance}d (left panel), where one can clearly see that the Antenna reshapes the emission into a main lobe oriented in the upward direction (dark lines), in contrast with the rather isotropic emission in the Reference case (light blue lines). The directivity enhancement at this wavelength is calculated to be $\left<\text{D}_\text{em}/\text{D}_\text{em}^0\right>_\text{exp}=1.43$ within $\text{NA}_\text{col}=0.3$ (see Eq.~(\ref{eq:directivity}) in Methods section for the formula used).
        
        We also computed the radiation patterns (using reciprocity, see Methods section for more details), shown in Fig.~\ref{fig4_enhance}d (right panel). One can see a good qualitative agreement between experiment and simulation, showing out-of-plane directivity. These simulation results give a simulated directivity enhancement of $\left<\text{D}_\text{em}/\text{D}_\text{em}^0\right>_\text{th}=1.31$ in the upward direction, which matches relatively well with the value extracted from the experimental measurements. 
        We thus deduced by applying the relation (non-rigorous) $\left<\eta_\text{em}\right>_\text{th}\approx\left<\text{D}_\text{em}/\text{D}_\text{em}^0\right>_\text{th}\left<\gamma_\text{r}/\gamma_\text{r}^0\right>_\text{th}$ that the theoretical average radiative decay rate enhancement (i.e. radiative yield) is $\left<\gamma_\text{r}/\gamma_\text{r}^0\right>_\text{th}=157.4$.

        \section{Discussion}
        
        The theoretical average enhancement factor $\left<\text{EF}_\text{th}\right> = 1263$ calculated from Eq.~(\ref{eq:ef_bis}) overestimates the experimental average enhancement factor $\left<\text{EF}_\text{exp}\right> = 654$. However, this can be expected if one remembers that Eq.~(\ref{eq:ef_bis}) used to estimate the theoretical enhancement factor is valid under the approximation that $\tau/\tau^0\approx 1$. Based on the experimentally obtained value from the measured lifetimes with a time-resolved PL setup (see Methods section), we obtain, after fitting the experimental data (see Supp. Info. Section~S7), a lifetime reduction of $\left<\tau/\tau^0\right>_\text{exp}^{-1}\approx 2.19$, that is $\left<\tau/\tau^0\right>_\text{exp}\approx 0.46$. Therefore, Eq.~(\ref{eq:ef_bis}) will overestimate the experimental enhancement factor of roughly a factor of two (non-rigorous), which we found to be the case. A more quantitative estimation of the theoretical average enhancement factor that takes into account the value of the intrinsic quantum yield $\text{QY}^0$ provides $\left<\text{EF}_\text{th}\right> = 790$, in better agreement with the experimental value (see Supp. Info. Section~S8). Note that another factor that may contribute to overestimate the experimental enhancement factor is that in theory we calculate it in a single direction, i.e. the upward direction, where the emission is maximum, while in the experiment we average over a $\text{NA}_\text{col}=0.3$ (chosen to average out noise present in the optical setup).
        
        We also analyse in Supp. Info. Section~S7 the results obtained for two other nanoantennas, called Antenna B and C (the one shown in the main text being called Antenna A in Supp. Info.), having approximately identical outer diameters and heights, but larger inner diameters, namely $D_\text{in}=80\,$nm and $D_\text{in}=110\,$nm, respectively. We found similarly that the theoretical calculations of the total enhancement factors overestimates the one found in experiment, by roughly the factor corresponding to the lifetime reduction. Moreover, this comparative study between Antennas A, B and C highlights that one can  tune (blue-shift) the resonance exploited for emission by varying (increasing) $D_\text{in}$ with respect to the resonance exploited for absorption, as already anticipated from the mode study in Fig.~\ref{fig1_designsalad}e. We also note that as the inner diameter $D_\text{in}$ increases, the total emission decreases, in agreement with the calculation of the modal Purcell factor which decreases as well (shown in Supp. Info. and which we recall quantify the coupling of the QDs with the strongly radiative antenna mode).
        
        Finally, in order to check if our design can accomodate different QDs sizes and types, we simulated the average emission enhancement factor $\left<\eta_\text{em}\right>_\text{th}$ for different nanogap sizes and refractive index of the spacer layer. The results (see Supp. Info. Section~S9) reveal that despite variation in terms of resonance strength, the spectral position of the resonances are not really affected, which was also reported in \cite{sugimoto2018broadband}, in constrast will all-plasmonic nanopatch antenna for which the resonances are more sensitive, making this hybrid system quite robust for a diversity of active materials.

	\section{Conclusion}

        In this work, we experimentally demonstrated an efficient and multi-resonance silicon nanoring on gold mirror nanoantenna coupled to localised quantum dots in a nanogap of $\sim 10\,$nm, which transformed very poor emitters (with an intrinsic quantum yield of $\text{QY}^0=3.8\times 10^{-3}$) into bright and directional sources. A total fluorescence enhancement factor of $654\times$ was measured, coming from different mechanisms that contributed to increase the brightness of our Antenna. We calculated in theory average enhancements of $\left<\gamma_\text{r}/\gamma_\text{r}^0\right>_\text{th}=157.4$, $\left<\gamma_\text{exc}/\gamma_\text{exc}^0\right>_\text{th}=7.3$ and $\left<\text{D}_\text{em}/\text{D}_\text{em}^0\right>_\text{th}=1.31$, for the radiative yield, excitation and directionality, respectively, and we experimentally obtained an average lifetime reduction of $\left<\tau/\tau_0\right>_\text{exp}^{-1}=2.19$. One can infer from this analysis that the radiative emission enhancement is the main mechanism responsible for the total fluorescence enhancement observed here, with also non-negligible contribution of the excitation enhancement, and modest collection efficiency gain and lifetime reduction.
        
        One particularity of the nanoring antenna is that it supports several resonances with strong scattering properties --- the antenna mode resonances to which the emitters are efficiently coupled --- that increase the radiative rates of emitters and reshape the directionality of emission within a main lobe pointing in the upward direction. These resonance wavelengths can be easily tuned using the ring \emph{inner diameter} to match with the emission wavelength of the emitters.
        Furthermore, the metallic mirror creates a nanogap that supports localized modes (gap mode resonances)\cite{filter2012circular,yang2017low}, whose resonance wavelength mostly depends on the ring \emph{outer diameter} and therefore can be easily tuned using this parameter, in an almost independent way from the resonances exploited for the emission enhancement. These were used to create intensity ``hot spots'' in the nanogap, leading to an overall gain in the excitation efficiency of the QDs.  
                
        Among the future improvements that should be carried out on these types of nanoantennas, we would like to mention a few here. First of all, there is a need to pay attention to ensure that the QDs are protected and not degraded by the fabrication process (i.e. not as reported in this work and in Ref.~\cite{sugimoto2018broadband}), in order to guarantee the integrity of the quantum emitters. Secondly, by exploiting lower order gap mode resonances, one could increase the excitation enhancement (by a factor 4 if one makes use of the second order according to our simulations --- not shown here), to potentially bring the total fluorescence to 3 orders of magnitude enhancement, figures reported so far only in all-plasmonic nanopatch antennas. It was not possible in this work because the second order gap mode could not be well separated spectrally from the emission of the QDs, and therefore we exploited a higher order gap mode (third order), which provided more modest excitation enhancement. 
        
        Among the advantages and potentials of such hybrid structures,  we would like to highlight that they could accommodate different QD sizes and could easily be enlarged or shrunk to shift the resonances to near-IR or UV wavelengths. Shifting to near-IR wavelengths would, in principle, be easier because of lower losses in the silicon at those wavelengths, while shifting to UV wavelengths might require using a different material with lower losses in the wavelength range for the ring structure.
 
        The experimental demonstration provided in this paper confirms the relevance of nanorings in hybrid dielectric-plasmonic nanostructures as highly tunable nanoantennas with subwavelength size, to create efficient, bright and directional photon sources in the visible spectral range, that can be of foremost importance for the next-generation of light-emitting devices.

	\section{Methods}
        
        \subsection{Fabrication}
        	To fabricate the nanoantenna structure, we deposited a $100\,$nm thick film of gold onto a silicon substrate with a $5$~nm titanium adhesion layer by Electron-beam Physical Vapor Deposition (EBPVD, Denton Explorer) at a rate of $0.1\text{\AA} \slash s$. 
        	
        	Next, we deposited a first layer of alumina (Al$_2$O$_3$) with thickness of $\approx3\,$nm on the gold using Atomic Layer Deposition (ALD, Beneq TFS 200) \cite{jin2019charge}, from trimethylaluminum and H$_2$O precursors at $120\degree$C. After that, a layer of CdSe/ZnS alloyed quantum dots, synthesized according to \cite{lim2014influence}, were spin-coated at $2000$~rpm for $1$~minute from a solution of $5$~mg QDs per ml in toluene. 
        	The quantum dots were then covered by another $\approx3$~nm thick layer of alumina, this time using ALD at a temperature of $80\degree$C. The final Al$_2$O$_3$/QDs/Al$_2$O$_3$ sandwich structure has a total thickness of approximately $10-15\,$nm (ellipsometry measurements estimated the thickness, assuming a homogeneous alumina layer, of $13\,$nm), prior to patterning the silicon (Si) ring nanoantenna.
        	
        	For the ring structure, we deposited a $230$~nm thick film of amorphous silicon by Induction-Coupled Plasma Chemical Vapor Deposition (ICP-CVD, Oxford PlasmaPro 100) at $80\degree$C from a SiH$_4$ precursor. Hydrogen silsesquioxane e-beam resist (Dow Corning XR-1541-06), spin coated at $5000$~rpm for $1$~minute and a change dissipation layer (Espacer 300AX01), spin coated at $1500$~rpm for $1$~minute were used for the Electron Beam Lithography (EBL) writing (Exlionix ELS-7000), with a dose of $\approx300$~mC/cm$^2$. The sample was then developed by a NaOH/NaCl salty solution ($1\%$~wt.~$/4\%$~wt. in de-ionized water) for $60$~s and then rinsed by de-ionized water to stop the development. The final structures was created by Induction-Coupled Plasma Reactive Ion Etching (ICP-RIE, Oxford Plasmalab 100) using chlorine gas, with a slight over etch to etch any quantum dots not protected by the silicon structures. A fabrication process flow schematics can be found in Supp. Info. Section~S10, Fig.~S14.

        	Since the fabrication process exerts thermal and chemical stress that can potentially degrade the QDs \cite{lyons2017addition}, we characterized optically the QD layer at all steps of the fabrication process to quantify the changes in photoluminescence (PL) and fluorescence lifetimes (see Supp. Info. Section~S10, Fig.~S15). While we observed that the quenching is successfully overcome by the presence of the first layer of Al$_2$O$_3$, the intensity of the PL is reduced by almost $98\%$ after depositing the second layer of Al$_2$O$_3$, that is in the configuration Au/Al$_2$O$_3$/QDs/Al$_2$O$_3$ (keeping the same excitation power). Moreover, the time-resolved PL experiments revealed that this drop in PL is correlated with a reduction of the QDs lifetime from $\tau_0=5.00\,$ns to $\tau_0=0.65\,$ns. From these observations, we estimated that the intrinsic quantum yield drops from $\text{QY}^0 = 0.29$ to $\text{QY}^0= 3.8\times 10^{-3}$, and concluded that the QDs are degrading because of thermal stress during the atomic layer deposition process used to deposit the second alumina layer \cite{lyons2017addition}.

        \subsection{Optical characterization}
            All optical measurements were performed in a microspectrometer setup, based on an inverted microscope (Nikon Ti-U) and a spectrometer system (Andor SR-303i spectrograph  with a $150$ lines/mm grating coupled to a $400\times1600$~pixel Andor Newton 971 EMCCD). Incident light was focused on the sample by a $100\times$ objective lens with a $0.9$NA (Nikon LU Plan Fluor). Signal collected by the same objective lens was then projected onto the spectrograph entrance slit with a width of $250\,\mu$m.
            \begin{itemize}
                \item \emph{Dark-field scattering measurements:} 
                    For dark-field scattering, white light from a halogen lamp was used to excite the sample, with the central low-$\vec{k}$ portion of the beam blocked from entering the objective lens, meaning only light scattered by the nanoantennas was collected and sent to the spectrograph. Reflectance of an silver mirror was used as the Reference.
                \item \emph{Photoluminescence spectroscopy:} 
                    For photoluminescence measurements, a supercontinuum source (SuperK Power, NKT Photonics) with band-pass filter (SuperK Varia, NKT Photonics), pulse duration $70$~ps, $78$~MHz repetition rate was used to excite PL. 
                    The band-pass filter was used to scan the pump wavelength from $488$ to $588$~nm with a $10$~nm bandwidth. Average pump power was maintained at $\approx250\,\mu$W. The pump laser was focused onto the sample substrate by the same $100\times$ $0.9$~NA objective lens, resulting in an approximately $1.37\,\mu$m diameter laser spot (see Supp. Info. Section~S6 for details on the method used to estimate the laser spot size). A $610$~nm long pass filter was used to cut off any pump laser light in the collection beam path, the $610$~nm cut-off can be clearly seen in all the photoluminescence curves in Supp. Info. Fig.~S1b. 
                    
                \item \emph{Back-focal-plane imaging:} 
                      To capture back-focal-plane images, the same $100\times$ $0.9$NA objective lens was used to collect light emitted by the nanoantennas, except that, instead of the image plane, the back focal plane of the objective was projected onto a CCD. The maximum collected angle, according to $NA = n \sin{\theta}$, and in our case, $n = 1$ (air), is about $\theta = 64.2\degree$. 
                
                \item \emph{Back-focal-plane spectroscopy:} 
                      To measure angle-resolved PL spectra, the same $100\times$ $0.9$NA objective lens was used to collect light emitted by the nanoantennas, except that, instead of the image plane, the back focal plane of the objective was projected onto the spectrograph entrance slit. The maximum collected angle, according to $NA = n \sin{\theta}$, and in our case, $n = 1$ (air), is about $\theta = 64.2\degree$. 
                
                \item \emph{Lifetime measurements:} 
                    Time-resolved photoluminescence was studied using a Picoquant Microtime 200 TCSPC system coupled to our microspectometer setup. The same supercontinuum source was used to excite the sample. Spectrally integrated PL in a narrow  $5$~nm range, centered at $650$~nm was collected using a Si single photon avalanche photodiode. The instrument response function (IRF) was recorded using excitation light scattered from the sample, where the IRF was measured to be $77$~ps. PL decay measurements were fit using reconvolution with the measured IRF by a bi-exponential function\cite{pavel_dmitriev_2022_6198822}.
            \end{itemize}
        
        \subsection{Numerical simulations}
            \begin{itemize}
                                        
                \item \emph{Quasi-normal mode calculations:} 
                    The quasi-normal modes (QNMs), denoted $\mathbf{E}_\alpha(\mathbf{r})$, can be defined as an eigenvalue problem of the solution of Maxwell equations in the absence of sources:
                    \begin{equation}
                        \nabla\times\frac{1}{\mu_0}\nabla\times\mathbf{E}_\alpha(\mathbf{r})=\omega^2_\alpha\varepsilon(\mathbf{r},\omega_\alpha)\mathbf{E}_\alpha(\mathbf{r})
                        \label{eq:method_Helmotz}
                    \end{equation}
                    where $\omega_\alpha=\omega_\alpha'+\mathrm{i}\,\omega_\alpha''$ denotes the complex eigenfrequency associated with the eigenmode $\mathbf{E}_\alpha(\mathbf{r})$, and supplemented by outgoing boundary conditions (also known as the Sommerfeld radiation condition as $|\mathbf{r}|\rightarrow\infty$). Note that $\omega_\alpha''<0$ due to the convention ``$\text{e}^{-\mathrm{i}\omega t}$'' used for the time-harmonic fields. Here, the system is considered nonmagnetic with a vacuum permeability $\mu_0$, and $\varepsilon(\mathbf{r},\omega)$ denotes the relative permeability of the medium. the complex eigenwavelengths are defined as $\lambda_\alpha\equiv 2\pi c/\omega_\alpha$. 
            
                    In order to solve Eq.~(\ref{eq:method_Helmotz}) and obtain the eigenmodes and eigenfrequencies in the configurations shown in Fig.~\ref{fig1_designsalad}e, we employed the ``QNMEig solver'', developped by IOGS-CNRS\cite{yan2018rigorous}, which, computes and normalizes the QNMs of plasmonic and photonic resonators, implemented using COMSOL Multiphysics. The QNMEig solver needs all dispersive material permittivities to be modelled by a $N$-pole Lorentz-Drude model, in order to reformulate Eq.~(\ref{eq:method_Helmotz}) into a linear eigenvalue problem (see, for example, \citet{yan2018rigorous}). 
                    The parameters of the Lorentz-Drude model that we used for the dispersive permittivities of amorphous silicon (nanoantenna) and gold (substrate) can be found in Supp. Info. Fig.~S11.

                     \item \emph{Near-field simulations:}
                    To compute the near-field, we used the finite-difference time-domain (FDTD) method in Ansys Lumerical FDTD. The Si particle (nanoantenna) in the presence of Au mirror was surrounded by a Total-Field Scattered-Field (TFSF) source, which simulates a plane wave excitation. The distance between the TFSF box and nanoantenna was set to be larger than $100\,$nm. The incident wave was chosen to be linearly polarized and coming at normal incidence from the top of the nanoantenna. The norm of the electric field was recorded in the plane located in the middle of the gap between the silicon nanoring and the gold mirror and then spatially averaged. We considered that the gap was filled with a homogeneous medium of refractive index corresponding to the one of alumina, i.e. $n=1.77$.

                \item \emph{Excitation enhancement simulations:}
                
                To compute the term $\eta_\text{exc}(\mathbf{r},\lambda_\text{exc})$ in Eq.~(\ref{eq:ef_bis}) shown in Fig.~\ref{fig3_gapfield}, we use the fact that the excitation rate is proportional to the local intensity of the electric field and therefore $\eta_\text{exc}(\mathbf{r},\lambda_\text{exc})$ can be readily expressed as\cite{chen2016manipulation}: 
        \begin{equation}
            \eta_\text{exc}(\mathbf{r},\lambda_\text{exc})=\left |\frac{\mathbf{u}\cdot\mathbf{E}(\mathbf{r},\lambda_\text{exc})}{\mathbf{u}\cdot\mathbf{E}_0(\mathbf{r},\lambda_\text{exc})}\right |^2
            \label{eq:excitation_enhancement}
        \end{equation}
        where we recall that $\mathbf{u}$ is a unit vector showing the orientation of the emitter dipole moment, $\mathbf{r}$ is the position of the emitter, and $\mathbf{E}(\mathbf{r})$ (resp. $\mathbf{E}_0(\mathbf{r})$) is the electric field at the emitter position $\mathbf{r}$ for a given excitation source in the nanoantenna case (resp. in the Reference case). We computed, using Ansys Lumerical FDTD, and considering as incident illumination a \emph{linearly} polarized planewave coming at \emph{normal incidence} from the top of the nanoantenna, the electric field intensity in the horizontal plane (i.e. parallel to the substrate) and located in the middle of the nanogap over an area corresponding to the ring horizontal cross-section. The averaged excitation enhancement  $\left<\eta_\text{exc}\right >$ was obtained after averaging over all directions to account for randomly distributed and oriented dipole emitters, and normalizing by the case without Si nanoantenna.
            
                \item \emph{Emission enhancement simulations:} 
                    The computation of the radiative emission enhancement $\eta_\text{em}(\mathbf{r},\lambda_\text{em})$ from Eq.~(\ref{eq:ef_bis}) and shown in Fig.~\ref{fig4_enhance} were carried out using the reciprocity principle\cite{PhysRevA.62.012712}, following the method well described in Ref.~\cite{zhang2015calculation}. This method was implemented in Ansys Lumerical FDTD, where planewave sources were used with two orthogonal linear polarizations and at normal incidence to excite the nanoantenna, using the TFSF source tool. Then, the near-field response was recorded in a plane located in the middle of the gap by point monitors, distributed in an area with the same size as the nanoantenna cross-section, and with a density of $3\,600\,\mu\text{m}^{-2}$ (i.e. we use approximately $400$ point monitors homogeneously distributed below the ring within an area of $\pi R^2$ with $R=D_\text{out}/2=190\,$nm). By reciprocity, the power recorded in each point monitor and calculated from the projection of the electric field along axis $i$ ($i=x$, $y$ or $z$) is equal to the emission power of light with the same polarization as the source from a point electric dipole oriented along $i$ and located at the same position as the monitor. To obtain the averaged emission enhancement $\left<\eta_\text{em}\right>$ of the assembly of electric dipoles randomly oriented and distributed uniformly under the nanoantenna, the power over all orientations $i$ and over the spatial distribution of monitors was integrated, averaged over two orthogonal linear polarizations, and normalized to the case without nanoantenna.

                \item \emph{Directional enhancement simulations:}
                    
                To obtain the directivity patterns shown in Fig.~\ref{fig4_enhance}d, we use the same reciprocal simulations as for the emission enhancement simulations, and made a sweep over all angles of incidence. In Fig.~\ref{fig4_enhance}d, we show the emission angular power distribution (averaged over two orthogonal linear polarizations). 
                
                In order to quantify the percentage of light that can be collected in the upward direction with given collection NA $\text{NA}_\text{col}$, we use the relation (see e.g. \cite{barreda2021metal}):
            \begin{equation}
                \text{D}_\text{em}=\frac{\int_{0}^{2\pi}\,\int_0^{\theta_\text{col}}p(\theta,\phi)\,\text{sin}(\theta)\text{d}\theta\text{d}\phi}{\int_{0}^{2\pi}\int_0^{\pi}p(\theta,\phi)\,\text{sin}(\theta)\text{d}\theta\text{d}\phi}
                \label{eq:directivity}
            \end{equation}
        with $p(\theta,\phi)$ being the angular power radiated into a certain solid angle (parametrized by $\theta$ and $\phi$), and $\theta_\text{col}$ is defined as $\theta_\text{col}=\text{sin}^{-1}(\text{NA}_\text{col})$. We consider only a single direction for the calculated radiation patterns, which is the upward direction, to be consistent with the enhancement spectra shown in Fig.~\ref{fig4_enhance}b, that is we set $\theta_\text{col}=0$ in Eq.~(\ref{eq:directivity}) (corresponding strictly speaking to $\text{NA}_\text{col}=0.0$).
        For the experimental directivity measurements, we choose to perform the integration over $\text{NA}_\text{col}=0.3$ in order to average the noise present in the collection channel, as explained in the main text.

            \end{itemize}
    	
	\begin{acknowledgement}
        The authors would like to thank Sergey Gorelik and Zhaogang Dong for helpful discussions regarding the fluorescence lifetime measurements. P.D. would also like to thank Jinfa Ho for assistance with nanofabrication. E.L. would like to acknowledge Christophe Sauvan for valuable advise regarding the quasi-normal mode analysis.
		The authors would like to acknowledge the Facility for Analysis, Characterisation, Testing and Simulation, Nanyang Technological University, Singapore, which provided their electron microscopy facilities. 
        Finally, we would like to thank the anonymous reviewers for their greatly detailed reviews which would have definitely contributed to bring significant improvements to our manuscript clarity and quality.
	\end{acknowledgement}

        \section{Funding Sources}
            This work was supported by the A*STAR SERC Pharos program (grant number 1527300025) and MTC Programmatic Grant No. M21J9b0085.
 
	\section{Author contributions}
	P.D., E.L., R.P.-D., H.V.D. and A.I.K. developed the concept.
	R.P.-D., J.K.W.Y., H.V.D. and A.I.K supervised and coordinated the work.
P.D. fabricated all the nanostructures and performed the SEM and all the optical measurements of the fabricated structures.
E.L. designed the nanoantenna and developed the code based on reciprocity principle used for the emission calculations and performed the numerical simulations of the near-field color maps, radiative emission spectra and emission angular patterns. E.L. also performed the quasi-normal mode analysis. P.D. also performed the calculations of the field enhancement in excitation configuration as well as some emission spectra and emission patterns calculations. 
L.D. provided support and supervised some of the optical measurements.
Z.D. developed part of the fabrication process and supervised some of the fabrication.
D.C.J.N. synthesized the QDs and helped to spin coat them onto the samples.
V.V. helped with SEM measurements.
P.D. and E.L. wrote the manuscript, and all co-authors participated in results interpretation and read and reviewed the manuscript.

	\begin{suppinfo}
	    Quantum dot emission/absorption spectra and fluorescence model; Resonance dependence on disk height; Gap mode resonance; Simulations of total fluorescence enhancement; Angle-resolved PL spectra; Laser spot size estimation; Comparison between antennas A, B and C; Enhancement dependence on QD quantum yield; Enhancement dependence on gap size; Nanoantenna fabrication process flow; Optical constants of amorphous silicon and gold.
	\end{suppinfo}

    \clearpage

	\clearpage
	\bibliography{refs}   
 
\end{document}


Number of pages: 27;
Number of figures: 16;
Number of tables: 4

\clearpage
\section{Quantum dot emission/absorption spectra and fluorescence model}

\begin{figure*}[!ht]
    	\centering
    	\includegraphics[width=0.7\textwidth]{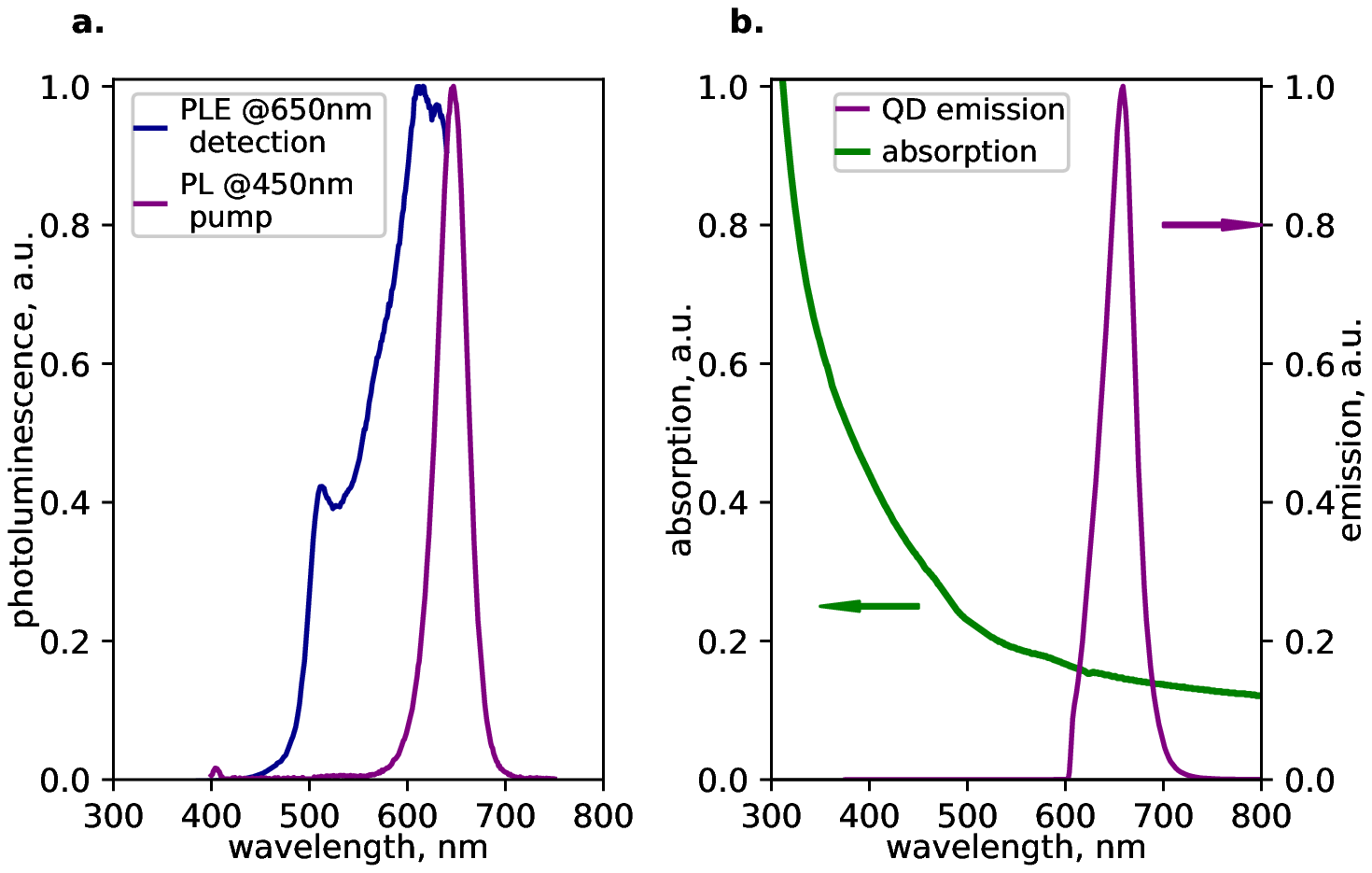}
    	\caption{\textbf{a.} Photoluminescence (PL) of the CdSe/ZnS quantum dots in solution (purple) with a $450\,$nm pump and photoluminescence efficiency (PLE) of the same QDs in solution (i.e. PL intensity at $650\,$nm detection wavelength depending on pump wavelength).
            \textbf{b}. Absorption (green) and emission (purple) spectra of the CdSe/ZnS quantum dots on top of glass (SiO$_2$). Pump laser wavelength was around $570\,$nm and a $610\,$nm long pass filter was used to cut off any pump laser light in the collection beam path.}
    	\label{qd_abs_em}
    \end{figure*}

Photoluminescence efficiency (i.e. photoluminescence intensity dependence on pump wavelength) and emission spectra for the QDs in solution are shown in Fig.~\ref{qd_abs_em}a. Experimentally measured absorption and emission spectra of the QDs deposited on a glass substrate are shown in Fig.~\ref{qd_abs_em}b.

\begin{figure*}[!ht]
    	\centering
    	\includegraphics[width=0.9\textwidth]{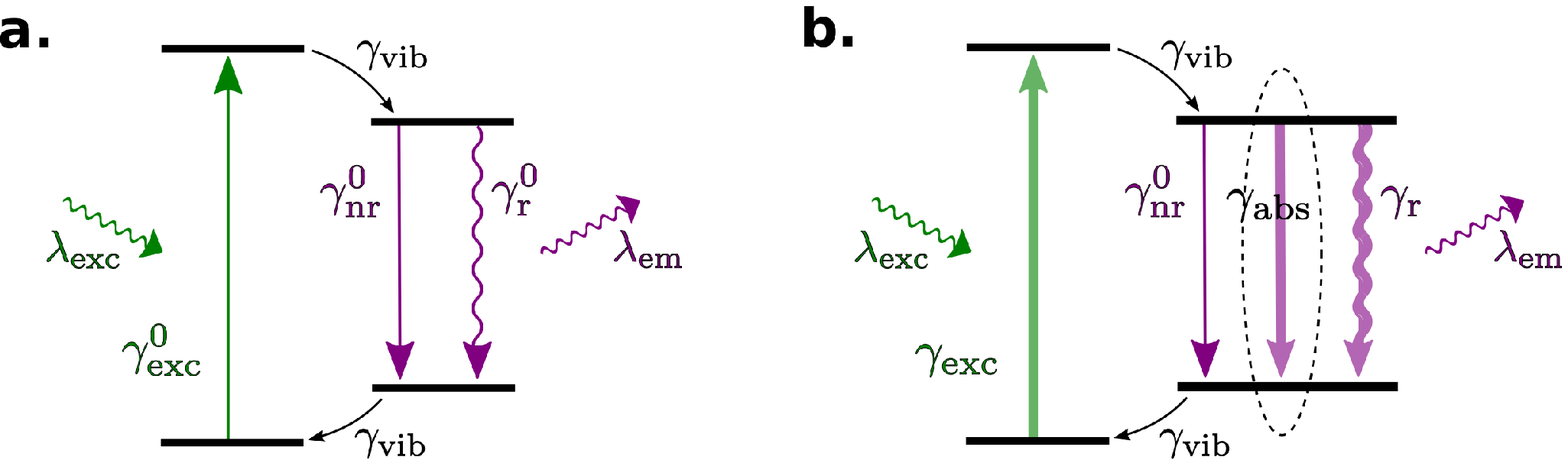}
    	\caption{Model of the energy levels of the QDs \textbf{a.} in the reference situation and \textbf{b.} in the nanoantenna situation.}
    	\label{fig:fluo}
    \end{figure*}

   The energy levels of the QDs in the reference situation (without a nanoantenna) are shown in Fig.~\ref{fig:fluo}a, and in Fig.~\ref{fig:fluo}b for a situation with a nanoantenna. 
   
\clearpage
\section{Resonance dependence on disk height}

    \begin{figure*}[!ht]
    	\centering
    	\includegraphics[width=0.7\textwidth]{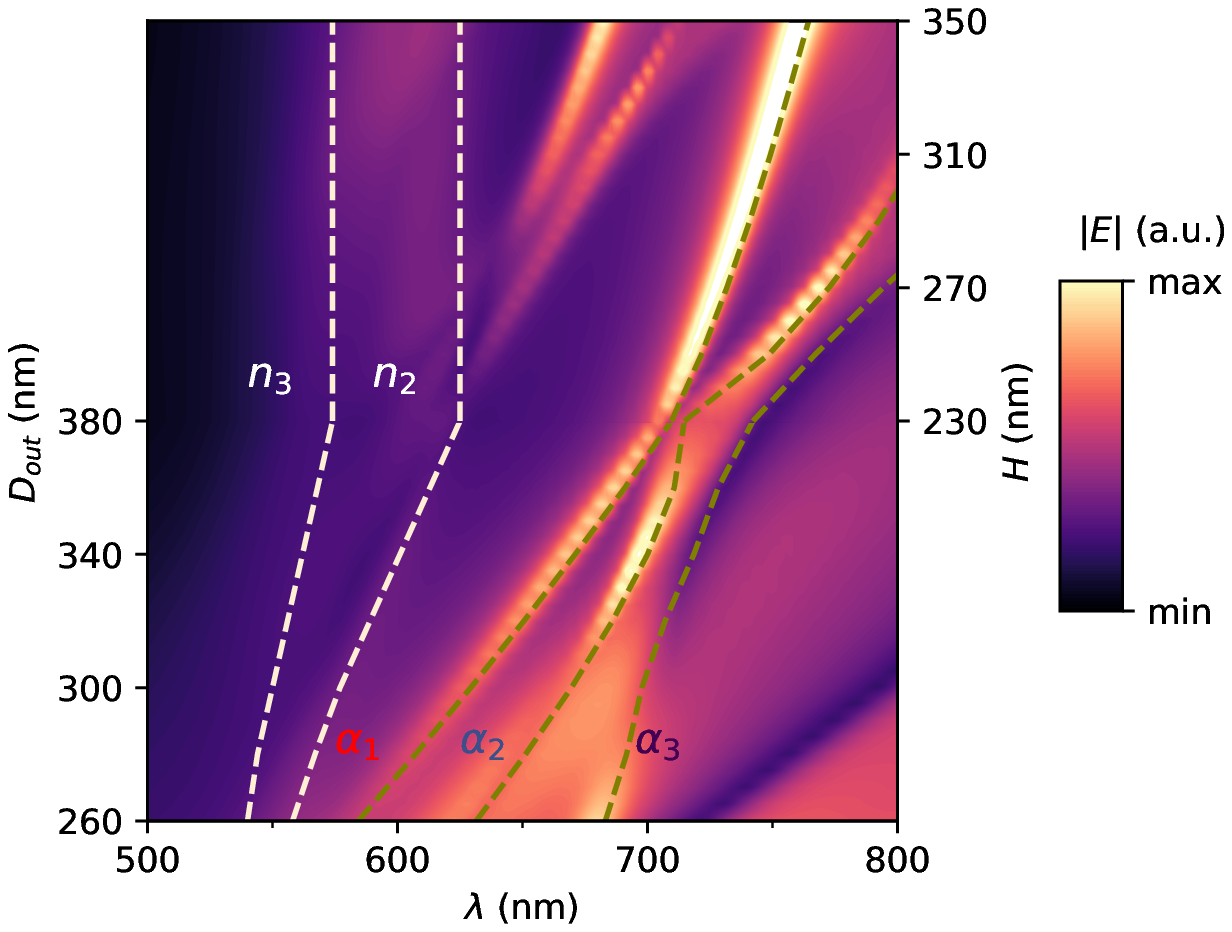}
    	\caption{Simulated electric field in nanoantenna gap, for $D_\text{in} = 0\,$nm and (lower half) increasing $D_\text{out}$ with fixed $H=230\,$nm or (upper half) increasing $H$ with fixed $D_\text{out}=380\,$nm.
    	Green dashed curves show calculated eigenwavelengths of three QNMs labelled $\alpha_1$, $\alpha_2$, $\alpha_3$. White dashed line show calcualted wavelengths of second and third order gap plasmon modes.}
    	\label{ring_height}
    \end{figure*}	

\clearpage

\section{Gap mode resonance}

\subsection{Multilayer dispersion relation of surface plasmon polariton}
    
    \begin{figure}[!ht]
    	\centering
    	\includegraphics[width=0.8\textwidth]{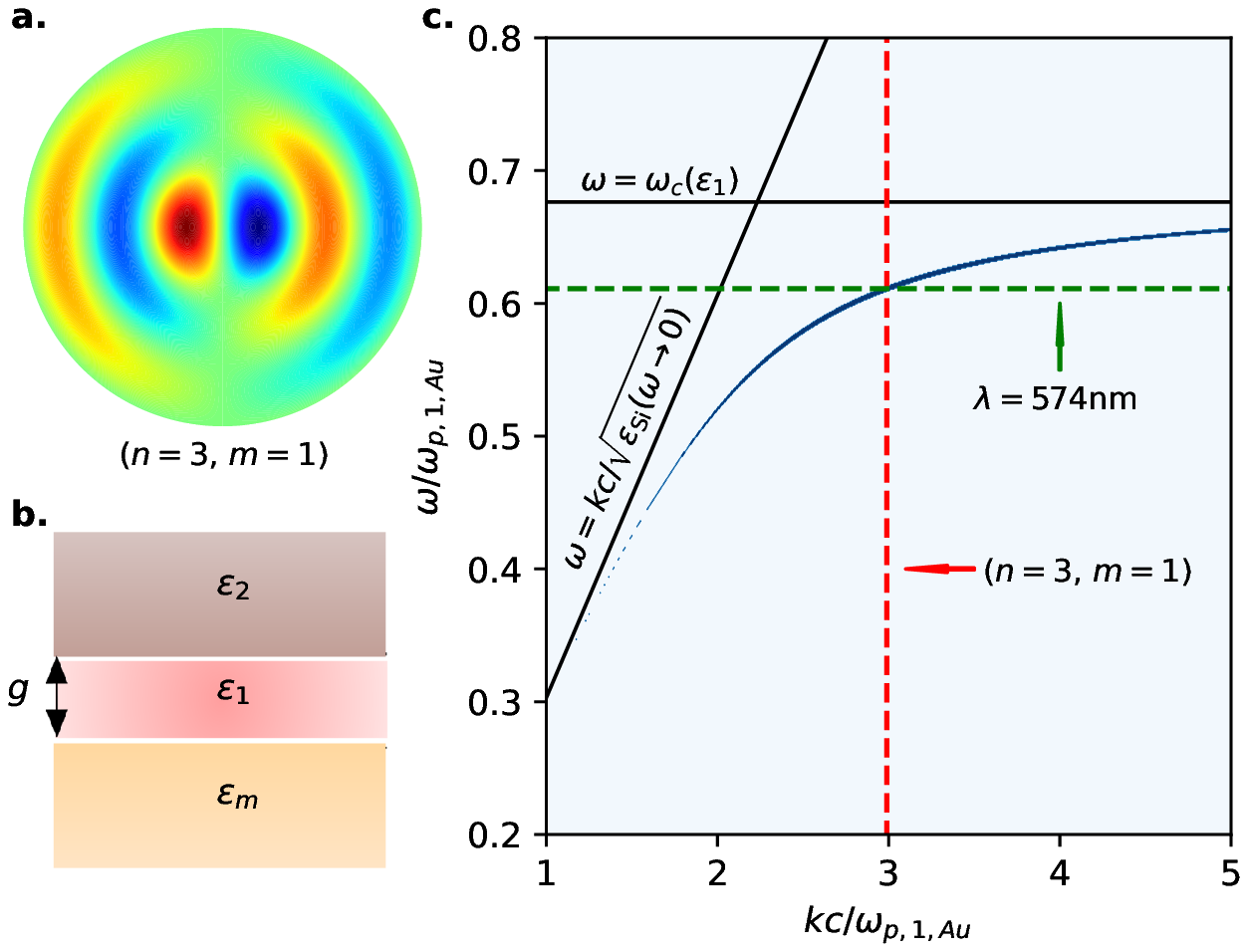}
    	\caption{
        \textbf{a.} Spatial distribution of the mode of a circular resonator of symmetry $(n=3,m=1)$, calculated from Eq.~(\ref{eq:mode_profile_theory}). Redder colours denote positive values, and bluer colour negative ones. \textbf{b.} Schematic of the metal/dielectric~1/dielectric~2 configuration.
    	\textbf{c.} Multilayer dispersion relation given by Eq.~(\ref{eq:multi_dispersion}) for a nanogap of thickness $g=10\,$nm and permittivity $\varepsilon_1=1.77^2$ (permittivity of Al$_2$O$_3$) sandwiched between an aSi semi-infinite covering layer of permittivity $\varepsilon_2=\varepsilon_\text{Si}\left(\omega\right)$ (given in Section~S11) without losses (i.e. $\gamma_{\text{Si}}=0$) and a semi-infinite gold mirror whose dispersive permittivity $\varepsilon_m=\varepsilon_\text{Au}(\omega)$ is the one (given in Section~S11) without losses (i.e. $\gamma_{1,\text{Au}}=0$ and $\gamma_{2,\text{Au}}=0$). The oblique dark line corresponds to the light cone in a homogeneous aSi medium $\omega=kc/\sqrt{\varepsilon_\text{Si}(\omega\rightarrow 0)}$ with $\sqrt{\varepsilon_\text{Si}(\omega\rightarrow 0)}=3.3$. The horizontal black line corresponds to the cutoff frequency $\omega_c(\varepsilon_1)$ obtained as the zero of the standard dispersion relation $k=\omega/c\sqrt{\varepsilon_1\varepsilon_m/(\varepsilon_1+\varepsilon_m)}$, that is by solving: $\varepsilon_1+\varepsilon_m(\omega_c)=0$.}
    	\label{multilayer_dispersion}
    \end{figure}
    

    The resonance frequencies (or wavelengths) associated to the quantized wavevectors are found by ``sampling'' the multilayer dispersion relation of surface plasmons in the configuration metal~/~dielectric~1~/~dielectric~2 according to \cite{yang2017low}, where in our case the metal is gold, dielectric~1 is alumina and dielectric~2 is silicon (configuration shown on Fig.~\ref{multilayer_dispersion}b, assuming the thickness of Si semi-infinite, which is justified if we are looking at the surface plasmon polariton localized in the gap). We thus calculate the multilayer dispersion relation given by solving the following transcendental equation \cite{karalis2005surface}:
    \begin{equation}
        \text{tanh}\left(k_1 g\right)=-\frac{1+\varepsilon_2 k_m/(\varepsilon_m k_2)}{\varepsilon_2 k_1/(\varepsilon_1 k_2)+\varepsilon_1 k_m/(\varepsilon_m k_1)}
        \label{eq:multi_dispersion}
    \end{equation}
    where ``tanh'' is the hyperbolic tangent function, $k_i=\sqrt{k^2-\omega^2/c^2\varepsilon_i}$, with $i=1,2$ for dielectrics 1 or 2 and $i=m$ for the metal substrate, and $g$ is the thickness of the dielectric 1 (the metal and dielectric 2 are considered semi-infinite). This dispersion is plotted in Fig.~\ref{multilayer_dispersion}c, in the case without losses (since it is easier to find the resonance frequencies as the dispersion curve is more sharply defined in the lossless case, and this assumption does not affect the position of the resonances).

\subsection{Mode profile}

The gap modes are modelled as the modes of a two dimensional circular resonator, which are found by solving the wave equation in polar coordinates $(r,\varphi)$ --- which becomes a Bessel's equation --- with Dirichlet boundary conditions\cite{ceperley_2016}. The solutions of this wave equation take the form \cite{ceperley_2016}:
    \begin{equation}
        E(r,\varphi,t)=A\cdot J_m(k_{mn}r)\cdot\text{cos}(m\varphi)\cdot\text{e}^{-\mathrm{i}\omega t}
        \label{eq:mode_profile_theory}
    \end{equation}
where $A$ is the amplitude of the mode, $k_{mn}$ is the quantized wavevector (according to Eq.~(3) of the main text), and $m$ is the azimuthal number. The spatial profile of the mode with symmetry $(n=3,m=1)$, calculated from Eq.~(\ref{eq:mode_profile_theory}) with $t=0$, is shown in Fig.~\ref{multilayer_dispersion}a, presenting a characteristic standing-wave pattern.

\begin{figure}[!ht]
	\centering
     \includegraphics[width=1.0\textwidth]{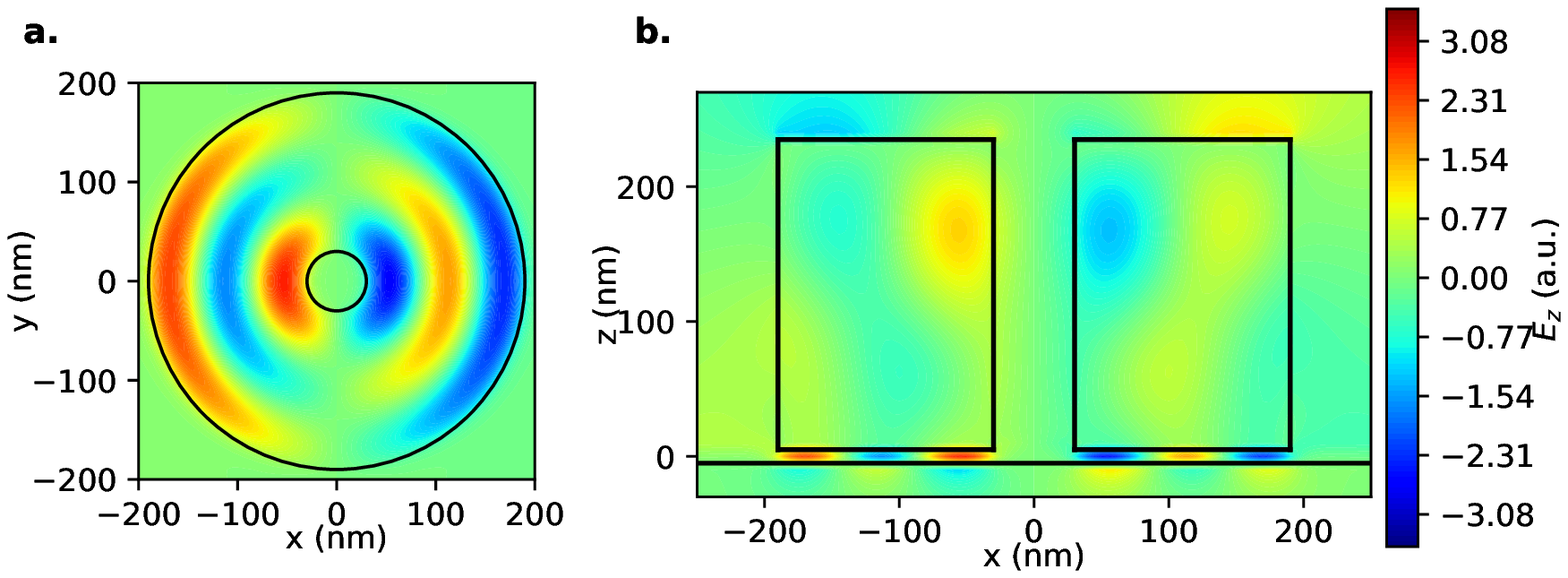}
	\caption{Component $E_z$ of the electric field at excitation wavelength $\lambda_\text{exc}=574\,$nm in (a) the horizontal cross-section $5\,$nm underneath the nanoring and (b) in the vertical cross-section passing through the middle of the nanoring, calculated with full-wave simulations.}
	\label{field_profiles}
\end{figure}

For comparison, we show full-wave simulations in Fig.~\ref{field_profiles} of the $z$-component of the electric field (the $x$- and $y$-components are negligible) corresponding to the same configuration as for Fig.~2 in the main text (which showed the local intensity enhancement). One can note the same symmetry as the (theoretically predicted) mode profile of a circular resonator shown in Fig.~\ref{multilayer_dispersion}a.

\clearpage
\section{Simulations of total fluorescence enhancement}
    \begin{figure*}[!ht]
    	\centering
    	\includegraphics[width=1.0\textwidth]{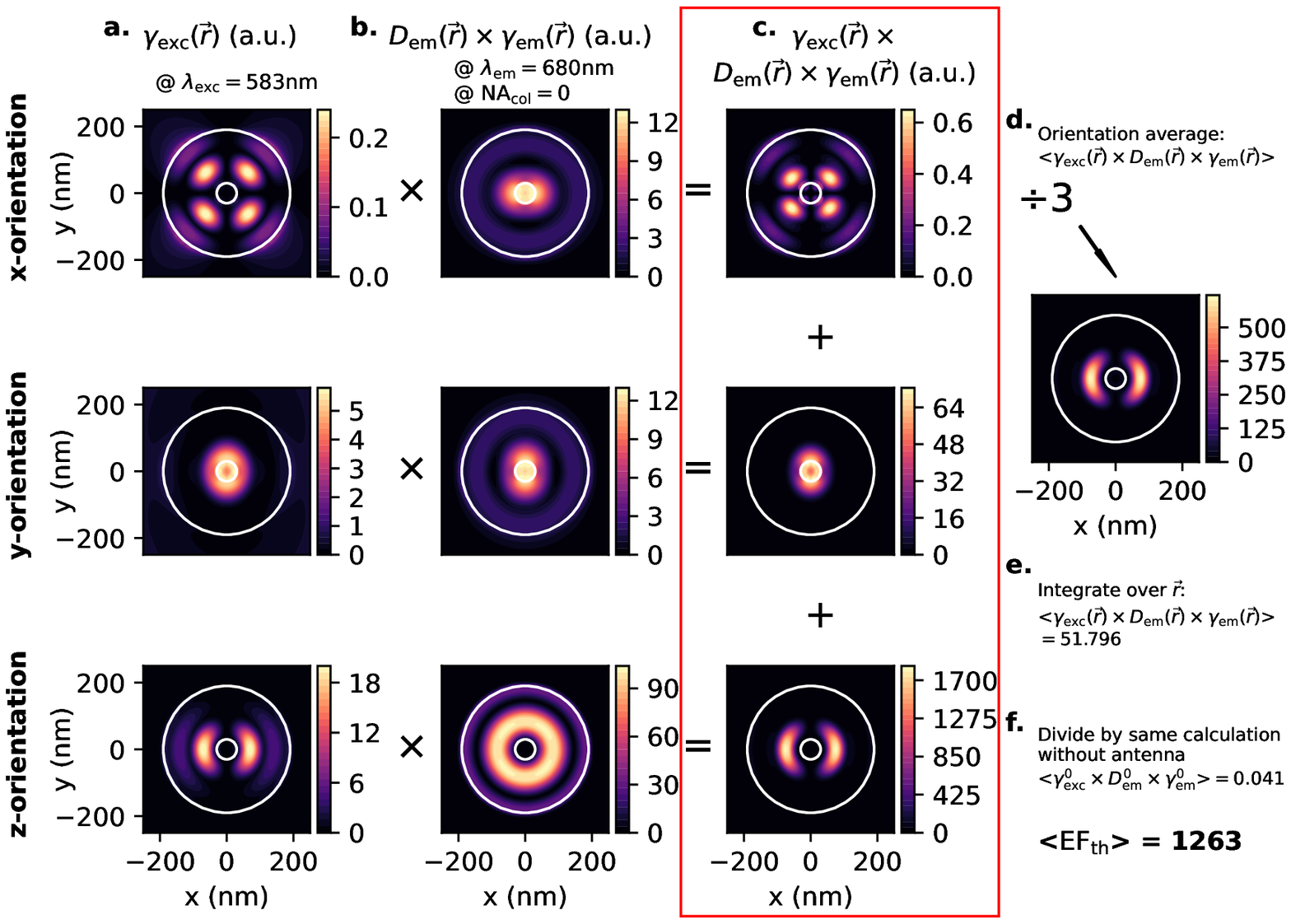}
    	\caption{Calculation of theoretical enhancement factor. \textbf{a.} Position-dependent $\gamma_\text{exc}(r)$ at gap mode resonance. \textbf{b.} Position-dependent $D_\text{em}(r) \times \gamma_\text{em}(r)$ at antenna resonance $\alpha_1$. \textbf{c.} Product of \textbf{a.} and \textbf{b.} showing enhancement provided by each excited field component. \textbf{d.} Averaging over the 3 orientations, assuming unpolarized emission. \textbf{e.} Averaging over the area of the nanoantenna. \textbf{f.} Normalization by the same calculation done in the absence of the nanoring structure, giving the theoretical value for <$\text{EF}_\text{th}$> of 1263.}
    	\label{rigorous_coupling}
    \end{figure*}

\clearpage
\section{Angle-resolved PL spectra}

\begin{figure*}[!ht]
    		\centering
    		\includegraphics[width=0.8\textwidth]{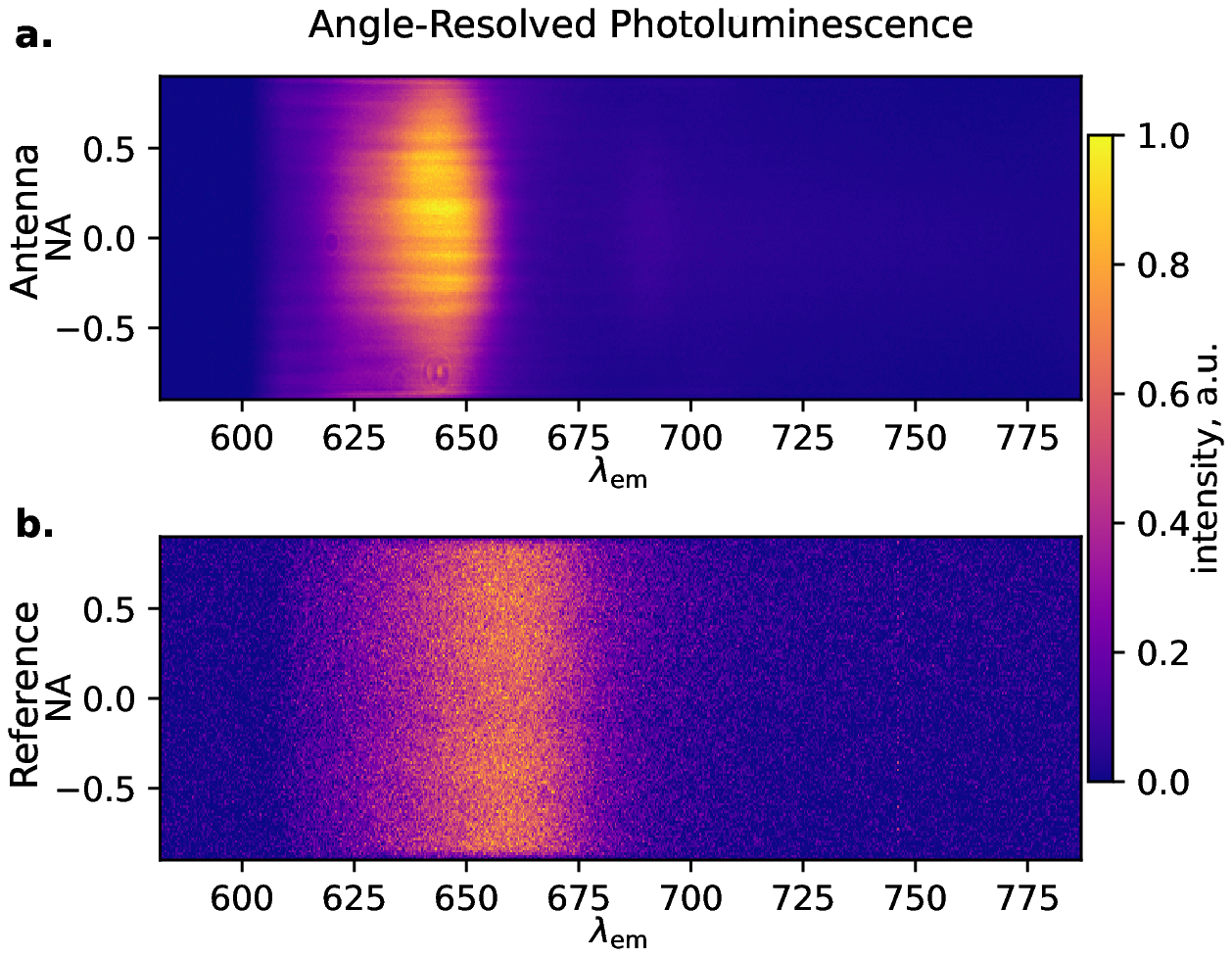}
    		\caption{Angle-resolved PL spectra in \textbf{a.} the Antenna case and \textbf{b.} the Reference case.
    		}
    		\label{fig3_arpl}
    	\end{figure*}

Experimental angle-resolved PL spectra of the Antenna were obtained by using an objective lens with $\text{NA} = 0.9$ and projecting the image onto a spectrometer. We show in Figs.~\ref{fig3_arpl}a and b the angle-resolved PL of the Antenna and Reference cases, respectively. The intensities are not normalized.

\clearpage
\section{Laser spot size estimation}

\begin{figure}[!ht]
	\centering
	\includegraphics[width=1.0\textwidth]{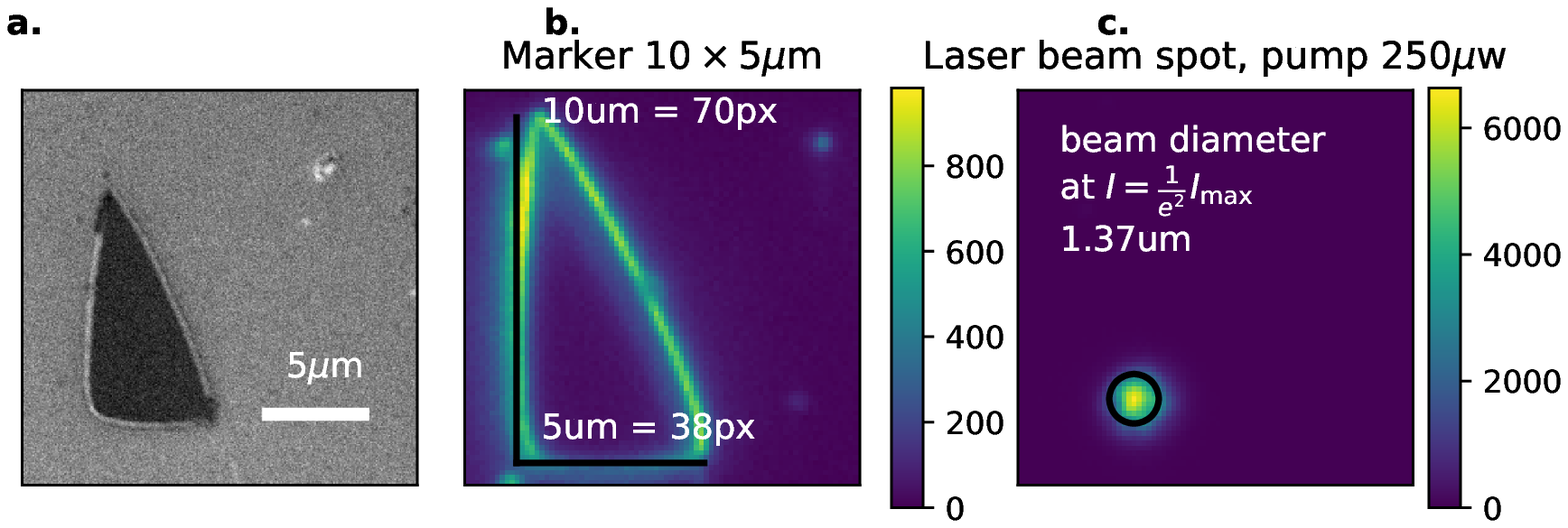}
	\caption{\textbf{a.} SEM of a marker on the sample, $10\times5\mu$m$^2$. \textbf{b.} Image of the same marker through $100\times$ $0.9$NA objective lens used to focus laser. Image scale is approximately $7$ pixels per micron. \textbf{c.} Image of the laser spot using the same optical setup. Measured spot diameter at intensity $I = \frac{1}{e^2}I_{max}$ is $1.37\mu$m.}
	\label{laser_spot}
\end{figure}

We estimated the size of our laser spot by comparing to a marker of well-known dimensions, as shown in Fig.~\ref{laser_spot}. The laser spot size was measured by exciting a uniform layer of quantum dots with the laser and capturing the image of the quantum dot emission, and then comparing that to an image of a marker captured through the same optical configuration. Spot diameter was estimated at intensity $I = \frac{1}{e^2}I_{max}$, giving $D = 1.37\mu$m.

\clearpage
\section{Comparison between antennas A, B and C}

In this Section, we compare three cases experimentally and in simulations of nanoantenna on top of gold mirror with same height $H=230\,$nm, outer diameter $D_\text{out}=380\,$nm, but different inner diameters: $D_\text{in}=60\,,80\,,110\,$nm, which we will call hereafter ``Antenna A'' (which is the Antenna discussed in the main text), ``Antenna B'' and ``Antenna C'', respectively.

\subsection{Emission enhancement spectra}

\begin{figure*}[!ht]
        	\centering
        	\includegraphics[width=0.9\textwidth]{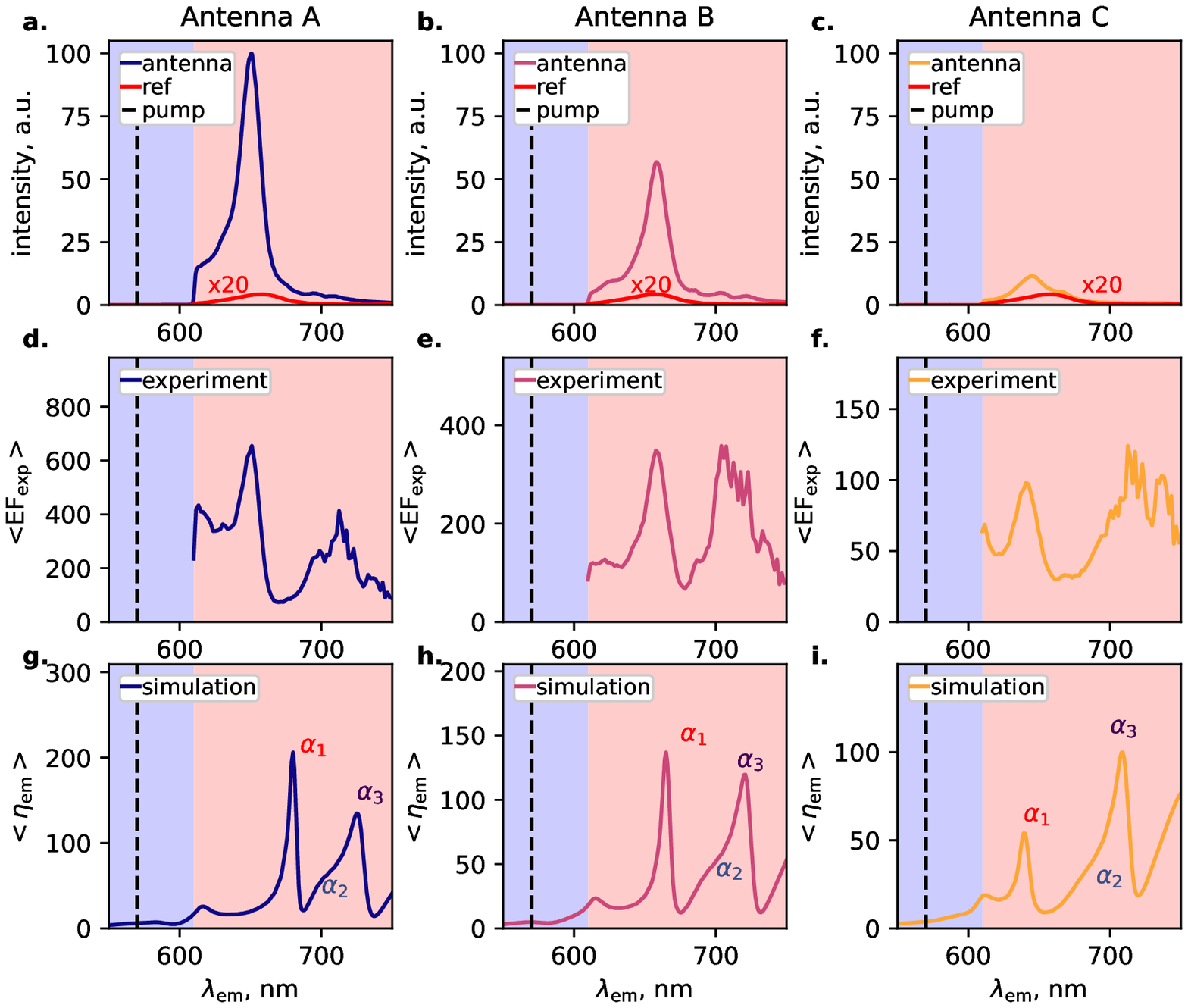}
        	\caption{\textbf{a, b, c} Experimental PL signal for the Antennas A, B and C (in blue, purple and orange, respectively) compared to Reference PL (in red). \textbf{d, e, f} PL enhancement factor $\left<\text{EF}_\text{exp}\right>$ obtained from dividing PL of Antennas A, B and C by Reference PL according to Eq.~(6) from the main text (same color code). \textbf{g, h, i} Simulated emission enhancement factor $\left<\eta_\text{em}\right>_\text{th}$ for the Antennas A, B and C (same color code). Dashed line in the plots indicates the pump wavelength. Blue area in the plots is experimentally filtered out in the collection channel, such that only signal from the red area reaches the spectrometer.}
        	\label{supp_enhance}
        \end{figure*}

We first show their measured experimental total enhancement spectra obtained for a collection $\text{NA}_\text{col}=0.3$, and compare them with simulations of the average emission enhancement $\left<\eta_\text{em}\right>$ in the normal direction in Fig.~\ref{supp_enhance}. As in the main text, photoluminescence enhancement was measured by dividing the photoluminescence intensity of each nanoantenna (normalized to the nanoantenna area) by the reference photoluminescence (layer of QDs without any nanoantenna, normalized to the laser spot area), see Eq.~(6) in the main text.

Experimentally, around the $650\,$nm, we have enhancement factors of $654$, $349$ and $98$ for Antenna A, B and C, respectively.
One can note that the positions of the peaks are slightly blue-shifted as one moves from Antenna B to Antenna C in experiment. Antenna A does not follow this trend, but this is attributed to slight variations in the height and outer diameter compared to the nominal values. 

In simulations, the peaks are blue-shifted as on goes from Antenna A to C as expected, and the emission wavelengths corresponding to the maximum of the first resonance are at at $680\,$, $665\,$ and $639\,$nm, respectively. The calculated emission enhancements $\left<\eta_\text{em}\right>$ in the normal direction are about $206$, $136$ and $54$, respectively.

\subsection{Angular radiation patterns}

\begin{figure*}[!ht]
    		\centering
    		\includegraphics[width=0.9\textwidth]{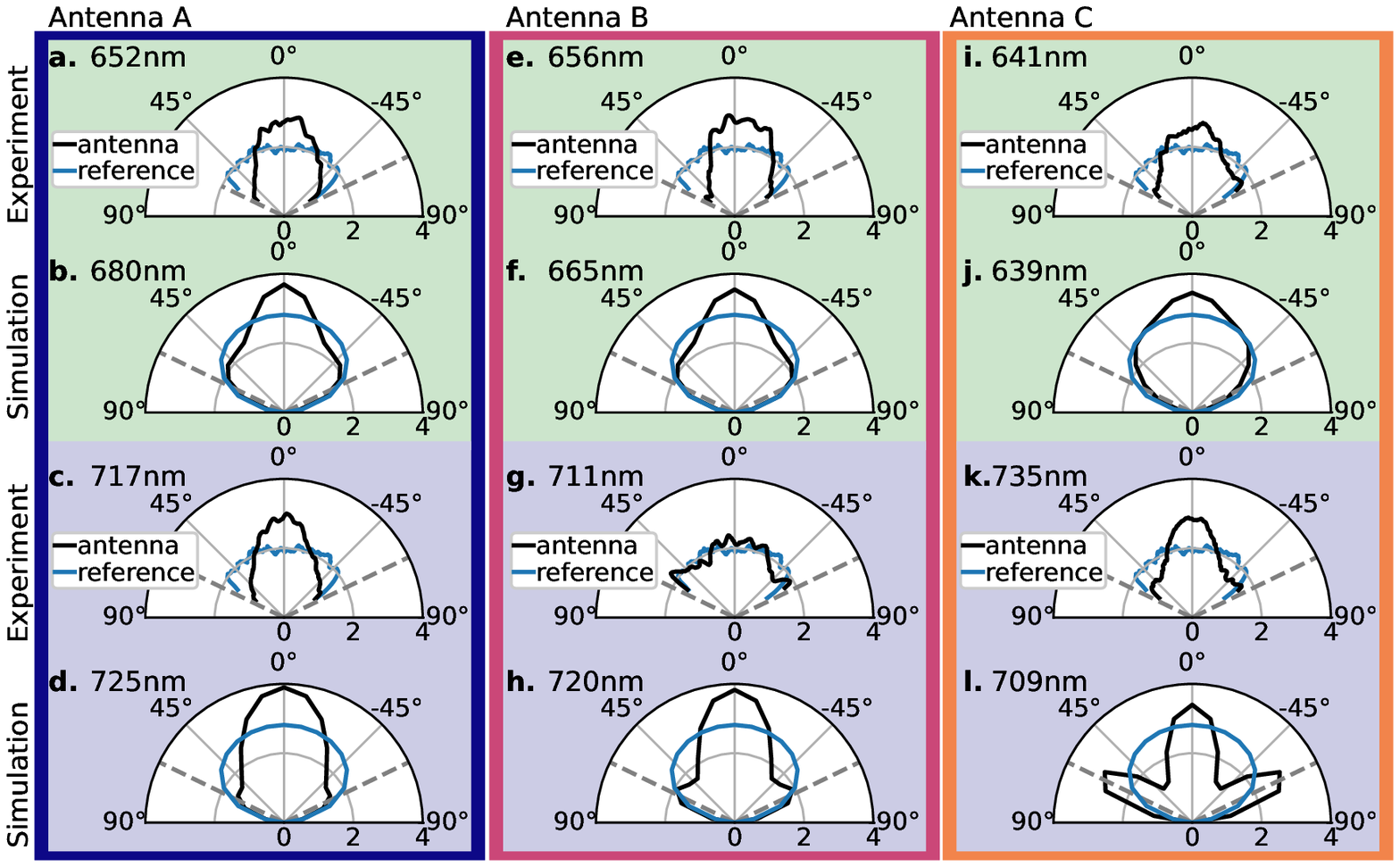}
    		\caption{\textbf{a, b, c, d} Experimental and simulated angular radiation patterns for the Antenna A (black curves) at the wavelengths --- displayed in the plots --- corresponding to the two main peaks ($\alpha_1$ and $\alpha_3$) in Fig.~\ref{supp_enhance}d (for experiment) and Fig.~\ref{supp_enhance}g (for simulations). The Reference cases are also shown both in experiment and simulation (light blue curves). \textbf{e, f, g, h} Same for Antenna B.  \textbf{i, j, k, l} Same for Antenna C. Grey dashed lines show the maximum collection angle accessible in our experiment.}
    		\label{supp_directivity}
    	\end{figure*}

Their radiation patterns are shown in Fig.~\ref{supp_directivity}, and reveal directivity in the out-of-plane direction in all cases.

\subsection{Lifetime measurements}
       
\begin{figure}[!ht]
    	\centering
    	\includegraphics[width=0.6\linewidth]{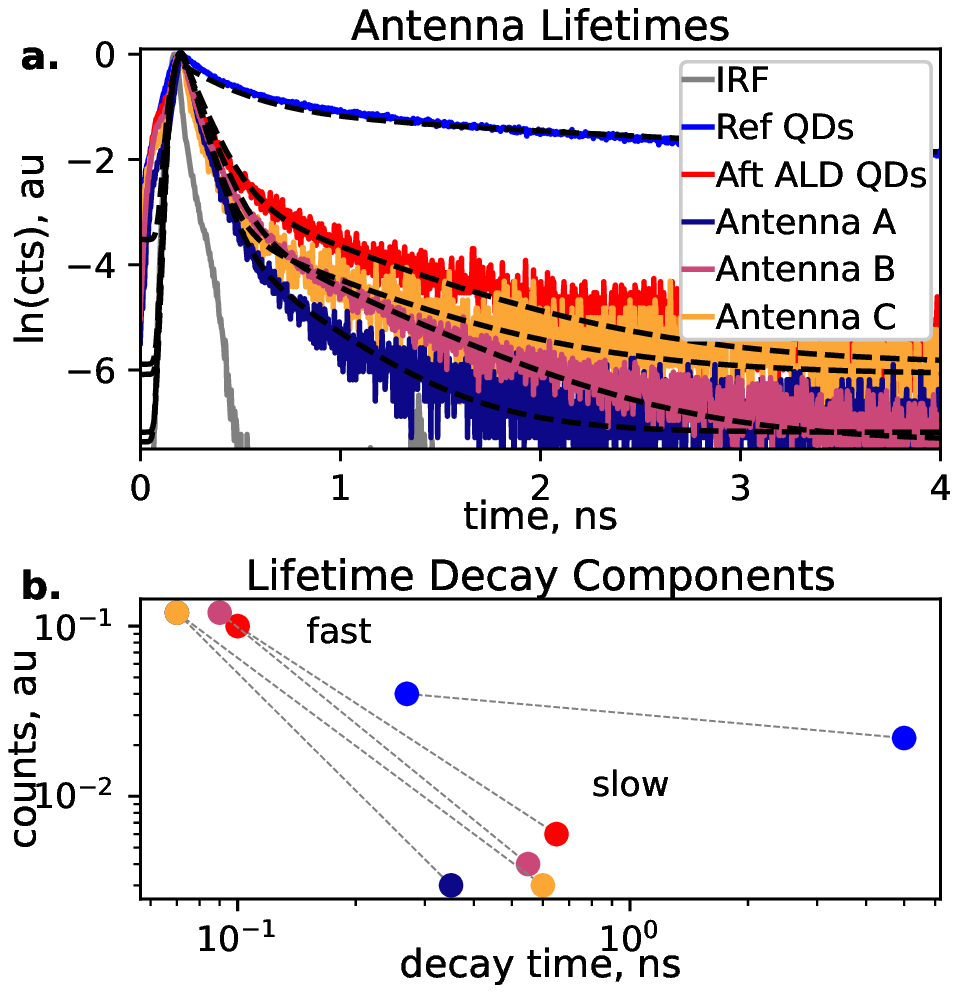}
    	\caption{\textbf{a.} Measured and fit lifetime signal from nanoantennas and reference quantum dots. Grey curve is the instrument response function. Blue curve is freshly spin-coated quantum dots lifetime. Red curve is the same quantum dots after ALD. Dark blue, violet and orange curves are the lifetime signals from Antennas A, B and C, respectively. Grey curve is the instrument response function. PL decay was measured for a narrow spectral band, centered at $650\,$nm. \textbf{b.} Decay components of lifetime signal fits.}
    	\label{lt_qy}
    \end{figure}

We also measure the lifetime for the nanoantennas A, B and C, which is shown in Fig.~\ref{lt_qy}, with the fitting parameters being displayed in Table~\ref{tab_lt_bis}, which each presents a ``fast'' decay and a ``slow'' decay. 

For each nanoantenna, weighting the relative contributions of each decay mechanisms is done be integrating the area under the fit components, which gives $R_1 = a_1 \times \tau_1$ and $R_2 = a_2 \times \tau_2$). This can then be used to get an ``average'' decay rate enhancement for each antenna relative to the reference quantum dots. As a result we get average decay rate enhancements for Antennas A, B and C of $2.19$, $1.35$ and $1.25$ , which corresponds to the averaged Purcell factor discussed in the main text. Note that this is the "total" Purcell factor, which is different from the "modal" Purcell factor calculated in Table~\ref{tab1} which is calculated for the antenna mode $\alpha_1$ only.
     
        \begin{table}[!ht]
        \caption{Amplitude ($a_1, a_2$) and decay ($\tau_1, \tau_2$) components of the bi-exponential fitting used to the the experimental lifetime measurements.}
        \centering	
        \begin{tabular}{ |p{5cm}|p{1.5cm}|p{1.5cm}|p{1.5cm}|p{1.5cm}|p{1.5cm}|p{1.5cm}|  }
            \hline
                & $a_1$, a.u. & $\tau_1$, ns & $R_1$, a.u. & $a_2$, a.u. & $\tau_2$, ns & $R_2$, a.u.\\
            \hline
            Antenna A   & $0.12$ & $0.07$ & $0.008$ & $0.004$ & $0.35$ & $0.001$\\
            \hline
            Antenna B   & $0.12$ & $0.09$ & $0.011$ & $0.004$ & $0.55$ & $0.002$\\
            \hline
            Antenna C   & $0.12$ & $0.07$ & $0.008$ & $0.003$ & $0.60$ & $0.002$\\
            \hline
        \end{tabular}
        \label{tab_lt_bis}
    \end{table}

\subsection{Modal Purcell factor computations}

In order to account for the different values of enhancement, we compute the Purcell factor $F_\alpha(\mathbf{r})$ associated to the antenna mode responsible for the peak around $650\,$nm (labelled $\alpha_1$ in the main text), according to Eq.~(1) in the main text. We found that the maximum Purcell factor $F_\alpha^\perp(\mathbf{r}^\text{max})$, corresponding to a dipole emitter with an out-of-plane orientation ($\perp$) and located at the position $\mathbf{r}^\text{max}$ where $\text{Re}(1/V_\alpha(\mathbf{\mathbf{r}^\text{max}}))$ is maximum (that is in the regions corresponding to the yellow/orange bands in Fig.~2b in the main text for the case of Antenna A), increases as the hole diameter decreases --- and is maximized for the disk case; see Table~\ref{tab1}. This indicates a stronger coupling of quantum emitters with the QNM as the hole diameter decreases. 
We also calculated the averaged Purcell factor denoted $\left<F_\alpha\right>$ for emitters spreading over the entire area below the ring (corresponding to the geometric cross-section of the ring $\sigma_\text{geo}=\pi R_\text{out}^2$ with $R_\text{out}=D_\text{out}/2=190\,$nm) and with random orientations; these values are displayed in Table~\ref{tab1} together with the quality factors, $Q_\alpha$. One can see that the averaged Purcell factor associated to this antenna mode decreases as the inner diameter increases, in agreement with the measured and simulated enhancement factors.

         \begin{table*}[!ht]
            \caption{Various quantities related to QNM $\alpha_1$ for four antennas configurations on gold mirror: Disk, Antenna A, Antenna B and Antenna C. Shown in the table are: the real and imaginary parts of the complex eigenfrequencies $\omega_\alpha$, the $Q$-factors associated with each mode calculated as $Q_\alpha=-\omega'_\alpha/(2\omega''_\alpha)$, the maximum Purcell factor associated to each mode calculated for a dipole with out-of-plane orientation and located in the positions where the coupling is maximum, and the averaged Purcell factor associated to each mode calculated after averaging over dipole orientations and emitter positions in the plane located below the antenna and in the middle of the gap.}
            \centering	
            \begin{tabular}{ |p{3cm}|p{3cm}|p{3cm}|p{1.5cm}|p{1.5cm}|p{1.5cm}|  }
                \hline
                & $\omega_\alpha'$ [rad/s] & $\omega_\alpha''$ [rad/s] & $Q_\alpha$ & $F_\alpha^\perp(\mathbf{r}^\text{max})$ & $\left<F_\alpha\right>$\\
                \hline
                Disk   & $2.654\times 10^{15}$ & $-4.402\times 10^{13}$ & $30.1$ & $41.6$ & $6.5$\\
                \hline
                Antenna A   & $2.748\times 10^{15}$ & $-4.849\times 10^{13}$ & $28.3$ & $27.4$ & $4.4$\\
                \hline
                Antenna B   & $2.811\times 10^{15}$ & $-5.188\times 10^{13}$ & $27.1$ & $22.0$ & $3.4$\\
                \hline
                Antenna C   & $2.923\times 10^{15}$ & $-5.438\times 10^{13}$ & $26.9$ & $14.1$ & $2.2$\\
                \hline
            \end{tabular}
            \label{tab1}
        \end{table*}

\subsection{Comparison simulation \emph{vs} experiment: summary}

We summarized the different calculated and measured enhancement factors in Table~\ref{tab}, and found good qualitative agreement between experiment and calculations.
    
    \begin{table*}[!ht]
            \caption{Comparison between calculated and measured averaged fluorescence enhancement factors. The ``theoretical'' enhancement factor $\left<\text{EF}_\text{th}\right>$ is calculated based on Eq.~(5) from the main text, where all the contributing ratios are computed theoretically. The ``experimental'' enhancement factor $\left<\text{EF}_\text{exp}\right>$ is obtained from the measured PL based on Eq.~(6) from the main text. All the quantities that depend on the excitation wavelength and emission wavelength are taken at around $\lambda_\text{exc}\approx 570\,$nm and $\lambda_\text{em}\approx 650\,$nm, respectively. The theoretical enhancement factor is calculated for a single direction (normal incidence), while experimentally we integrate the signal over $\text{NA}_\text{col}=0.3$. The other gains obtained either from simulations (with subscript "th") or experimentally (with subscript "exp") are also displayed.}
            \centering	
            \begin{tabular}{ |p{2.0cm}|p{2.0cm}|p{1.6cm}|p{2.2cm}|p{2.4cm}|p{1.6cm}|p{1.3cm}|p{1.4cm}|  }
                \hline
                & $\left<\gamma_\text{exc}/\gamma_\text{exc}^0\right>_\text{th}$ &
                $\left<\gamma_\text{r}/\gamma_\text{r}^0\right>_\text{th}$ &
                $\left<\text{D}_\text{em}/\text{D}_\text{em}^0\right>_\text{th}$ & $\left<\text{D}_\text{em}/\text{D}_\text{em}^0\right>_\text{exp}$ &  $\left<\tau/\tau_0\right>_\text{exp}^{-1}$ & $\left<\text{EF}_\text{th}\right>$ & $\left<\text{EF}_\text{exp}\right>$\\
                \hline
                Antenna A   & $7.3$ & $157.4$ & $1.31$ & $1.43$ & $2.19$ & $1263$ & $654$\\
                \hline
                Antenna B   & $5.0$ & $107.9$ & $1.26$ & $1.45$ &  $1.35$ & $891$ & $349$\\
                \hline
                Antenna C   & $3.8$ & $43.9$ & $1.21$ & $1.34$ & $1.25$ & $234$ & $98$\\
                \hline
            \end{tabular}
            \label{tab}
        \end{table*}

\clearpage
\section{Enhancement factor dependence on intrinsic quantum yield}
In \cite{aouani2011bright}, the authors derive an interesting expression of the enhancement factor that shows the dependence in terms of the intrinsic quantum yield:
        \begin{equation}
        \text{EF}_\text{th}(\mathbf{r})        =\eta_\text{exc}(\mathbf{r})\times\eta_\text{em}(\mathbf{r})\times \frac{1}{(1-\text{QY}^0)+\text{QY}^0[\gamma_\text{r}(\mathbf{r})+\gamma_\text{abs}(\mathbf{r})]/\gamma_\text{r}^0}
        \label{eq:ef_bis}
        \end{equation}
    
    For averaging over positions/orientations, and assuming that $\gamma_\text{abs}(\mathbf{r})\approx 0$, we use the formula (non-rigorous):
        \begin{equation}
        \left<\text{EF}_\text{th}\right>       \approx\left<\eta_\text{exc}\eta_\text{em}\right>\times \frac{1}{(1-\text{QY}^0)+\text{QY}^0\left<\gamma_\text{r}/\gamma_\text{r}^0\right>}
        \label{eq:ef_bisbis}
        \end{equation}
    
\begin{figure*}[!ht]
    	\centering
    	\includegraphics[width=0.7\textwidth]{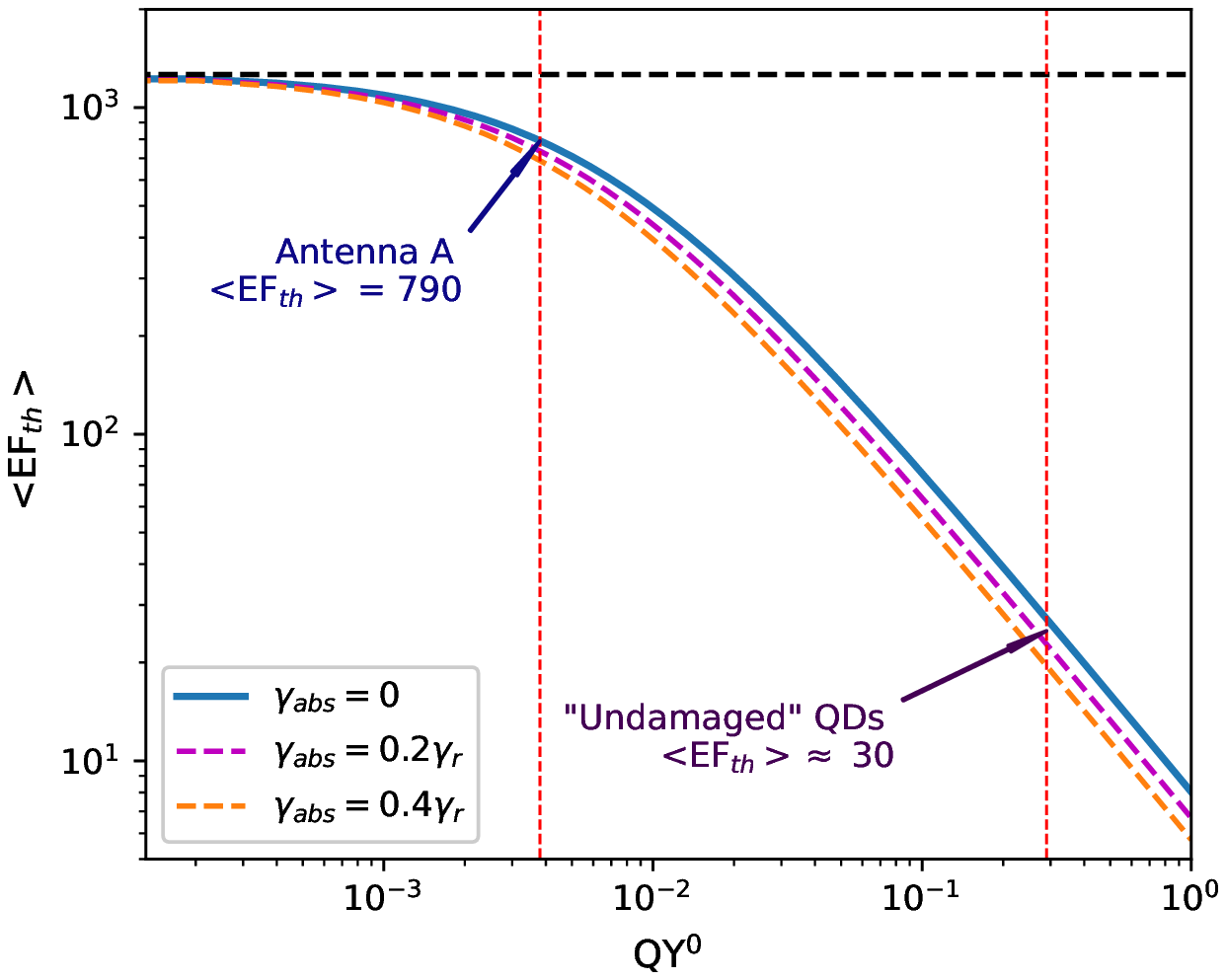}
    	\caption{Dependence of fluorescence enhancement factor on the intrinsic quantum yield $\text{QY}^0$ of the QDs. Horizontal black dashed line represents $\left<\text{EF}_\text{th}\right>=1263$. Vertical red dashed lines show experimental measured $\text{QY}^0$ --- before, $\text{QY}^0=0.29$, and after fabrication, $\text{QY}^0=3.8\times 10^{-3}$ (due to degradation). The horizontal and vertical axis are in logscale.}
    	\label{qy_effects}
    \end{figure*}

    In Fig.~\ref{qy_effects} we plot the theoretical average total enhancement $\left<\text{EF}_\text{th}\right>$ given by Eq.~(\ref{eq:ef_bisbis}) as a function of $\text{QY}^0$, using the values calculated in the main text for Antenna A, that is  $\left<\eta_\text{exc}\eta_\text{em}\right>_\text{th}=1263$ and $\left<\gamma_\text{r}/\gamma_\text{r}^0\right>_\text{th}=157.4$.
    From this, we extract at $\text{QY}^0=3.8\times 10^{-3}$ a value of total enhancement of $\left<\text{EF}_\text{th}\right>\sim 790$, which is closer to the experimental value $\left<\text{EF}_\text{exp}\right>=654$, compared to the value calculated in the main text based on Eq.~(5) of $\left<\text{EF}_\text{th}\right>\sim 1263$. The same can be done for the cases of Antenna B and Antenna C (not shown here), and we obtain the values $\left<\text{EF}_\text{th}\right>\sim 630$ and $\left<\text{EF}_\text{th}\right>\sim 200$, respectively, which are also in closer agreement with the experimental values reported in Table~\ref{tab}.
    
    Note that these calculated values may still overestimate since we neglected the absorption of the nanoantenna itself ($\gamma_\text{abs}(\mathbf{r})\approx 0$), but the effect of absorption would be to reduce slightly these values, as shown in Fig.~\ref{qy_effects}.

    \clearpage
\section{Emission enhancement dependence on gap size}
    \begin{figure*}[!ht]
    	\centering
    	\includegraphics[width=0.8\textwidth]{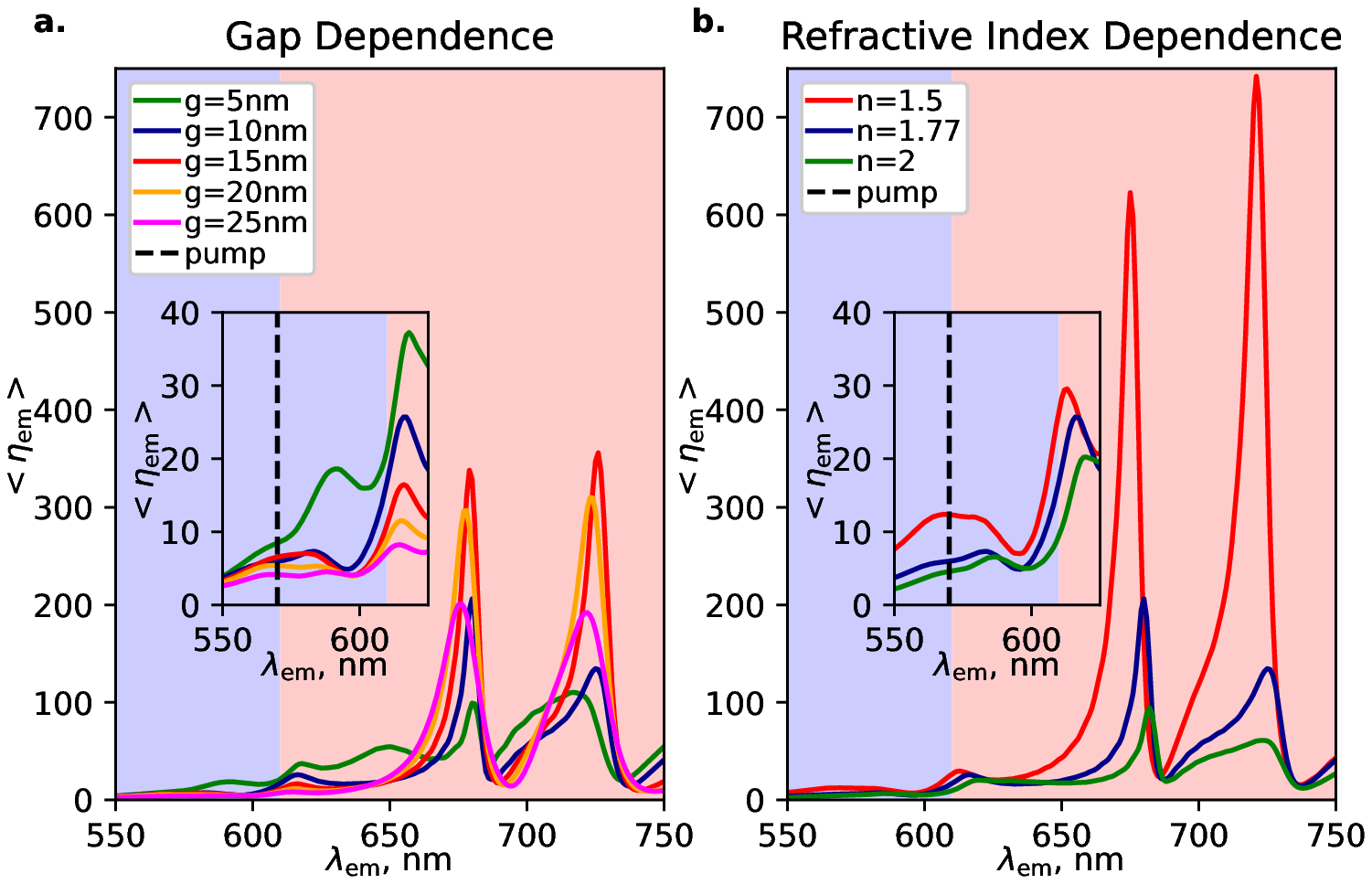}
    	\caption{Simulated enhancement factor $\left<\eta_\text{em}\right>_\text{th}$ for the Antenna discussed in main text for different gap sizes (\textbf{a}) or different gap refractive index (\textbf{b}).
    	Dashed line in the plots indicates the pump wavelength. Blue area in the plots would be filtered out in the collection channel such that only signal from the red area would reach the spectrometer. Insets show zoomed-in portions of plots.}
    	\label{gap_effects}
    \end{figure*}

\clearpage
\section{Nanoantenna fabrication process flow}

    \begin{figure*}[!ht]
        \centering
        \includegraphics[width=0.8\textwidth]{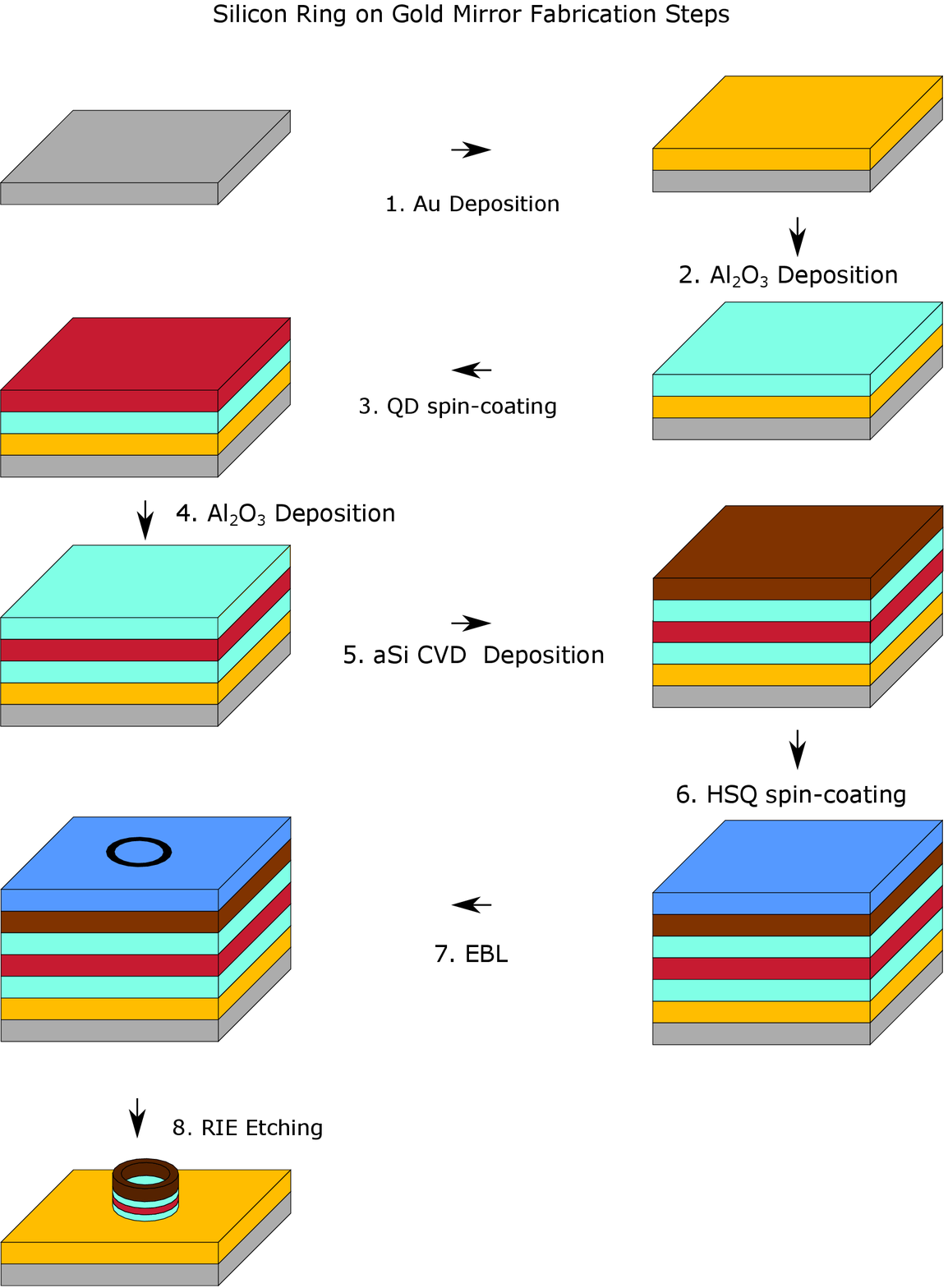}
        \caption{Schematics showing nanoantenna fabrication steps.}
        \label{fab}
    \end{figure*}
    The whole nanoantenna structure fabrication process flow is shown in Figure~\ref{fab}. Details of each step of the process are given below:

    \begin{enumerate}
        \item We deposited a $100\,$nm gold thin film onto a silicon substrate with a $5$~nm titanium adhesion layer by EBPVD (Denton Explorer) at a rate of $0.1\text{\AA} \slash s$.
        \item  Next, using ALD (Beneq TFS 200), we deposited several nanometers (30 cycles) of alumina from trimethylaluminum and H$_2$O precursors at $120\degree$C.
        \item After that, a layer of CdSe/ZnS alloyed quantum dots\cite{lim2014influence} were spin-coated at $2000$~rpm for $1$~min from a solution of $5\,$mg QDs per mL in toluene (shown on Figure~1c in the main text).
        \item The quantum dots were then covered by another layer of alumina (30 cycles), this time using ALD at a temperature of $80\degree$C.
        \item For the ring structure, we deposited an $\approx230\,$nm thick film of amorphous silicon by ICP-CVD (Oxford PlasmaPro 100) at $80\degree$C from a SiH$_4$ precursor (45 sccm SiH$_4$ and 30 sccm Ar at $8\,$mTorr process pressure, $50\,$W RF power at $20\,$DC forward bias and $3000\,$W ICP RF power).
        \item Hydrogen silsesquioxane e-beam resist (Dow Corning XR-1541-06), spin-coated at $5000$~rpm for $1$~min, and a change dissipation layer (Espacer 300AX01), spin coated at $1500$~rpm for $1$~min, were used for the EBL writing.
        \item EBL (Elionix ELS-7000) was performed at an acceleration voltage $100\,$kV and a current of $500\,$pA, with a dose of $\approx300\,$mC/cm$^2$. The sample was then developed in a NaOH/NaCl salty solution ($1\%$~wt.~$/4\%$ wt. in de-ionized water) for $60\,$s and then rinsed by de-ionized water to stop the development.
        \item  The final structure created by ICP-RIE etching (Oxford Plasmalab 100) using chlorine gas, with a slight over etch to etch any quantum dots not protected by the silicon structures ($22\,$sccm Cl$_2$ at $5\,$mTorr process pressure, $100\,$W RF power and $300\,$W ICP RF power).
    \end{enumerate}

\begin{figure}[!ht]
    	\centering
    	\includegraphics[width=1.0\textwidth]{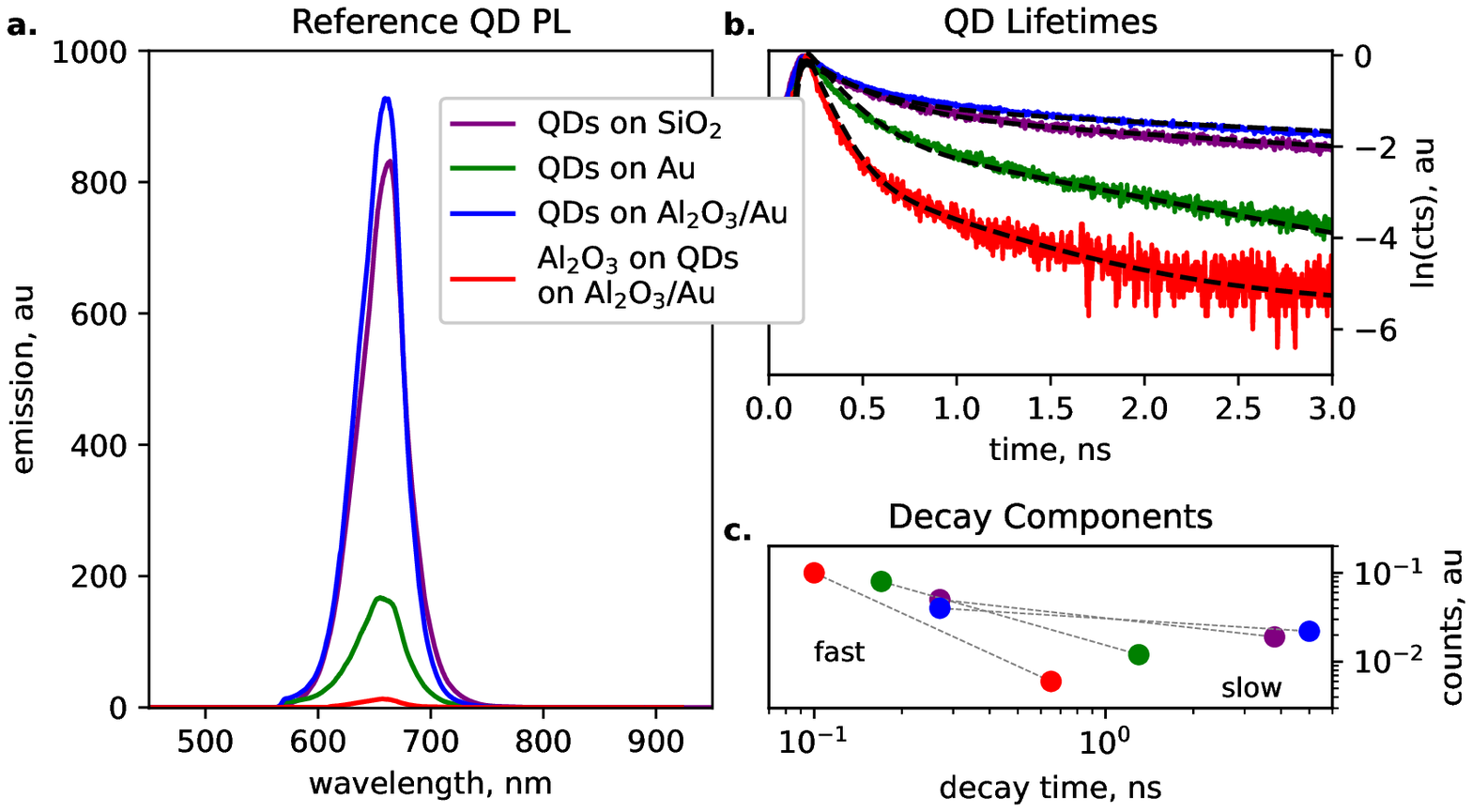}
    	\caption{Photoluminescence (\textbf{a.}), lifetimes (\textbf{b.}) and fit decay components (\textbf{c.}) of reference quantum dots. Purple lines - quantum dots on SiO$_2$. Green lines - quantum dots directly on Au mirror, with noticeable quenching. Blue lines - quantum dots on Au mirror but with Al$_2$O$_3$ spacer separating them from the gold, which completely stops quenching. Red lines - damaged quantum dots after second Al$_2$O$_3$ deposition.}
    	\label{ref_qds}
    \end{figure}

We characterized the QD emission by time-resolved and PL intensity measurements after fabrication steps 3 and 4, identified by the blue and red colors in Fig.~\ref{ref_qds}, respectively. For comparison, we also show the cases of QDs on a glass substrate and on a gold substrate, identified by the purple and green colors in Fig.~\ref{ref_qds}, respectively.

PL intensity measurements are shown in Fig.~\ref{ref_qds}a. Time-resolved PL decay shown in Fig.~\ref{ref_qds}b was fit by a bi-exponential function using reconvolution\cite{pavel_dmitriev_2022_6198822}. Fitting parameters are shown in Table~\ref{tab_lt}. 

For all measured samples, the time-resolved decay showed two characteristic decay rates, shown in Fig.~\ref{ref_qds}c, related to different decay mechanisms. These decay mechanisms can be related to interactions between the quantum dots in the QD layer\cite{guzelturk2014excitonics} or quantum dots with different orientations exhibiting different decay rates because of substrate interactions\cite{hoang2016ultrafast}. 

    \begin{table}[!ht]
        \caption{Amplitude ($a_1, a_2$) and decay ($\tau_1, \tau_2$) components of the bi-exponential fitting used to the experimental lifetime measurements.}
        \centering	
        \begin{tabular}{ |p{5cm}|p{1.5cm}|p{1.5cm}|p{1.5cm}|p{1.5cm}|p{1.5cm}|p{1.5cm}|  }
            \hline
            & $a_1$, a.u. & $\tau_1$, ns & $R_1$, a.u. & $a_2$, a.u. & $\tau_2$, ns & $R_2$, a.u.\\
            \hline
            QDs on SiO$_2$                         & $0.05$ & $0.27$ & $0.014$ & $0.019$ & $3.80$ & $0.072$ \\
            \hline
            QDs on Au                              & $0.08$ & $0.17$ & $0.014$ & $0.012$ & $1.30$ & $0.016$ \\
            \hline
            QDs on Al$_2$O$_3$/Au                  & $0.04$ & $0.27$ & $0.011$ & $0.022$ & $5.00$ & $0.110$ \\
            \hline
            Al$_2$O$_3$ on QDs on Al$_2$O$_3$/Au   & $0.10$ & $0.10$ & $0.010$ & $0.006$ & $0.65$ & $0.004$\\
            \hline
        \end{tabular}
        \label{tab_lt}
    \end{table}

\clearpage
\section{Optical constants of amorphous silicon and gold}

\begin{figure*}[!ht]
    	\centering
    	\includegraphics[width=1.0\textwidth]{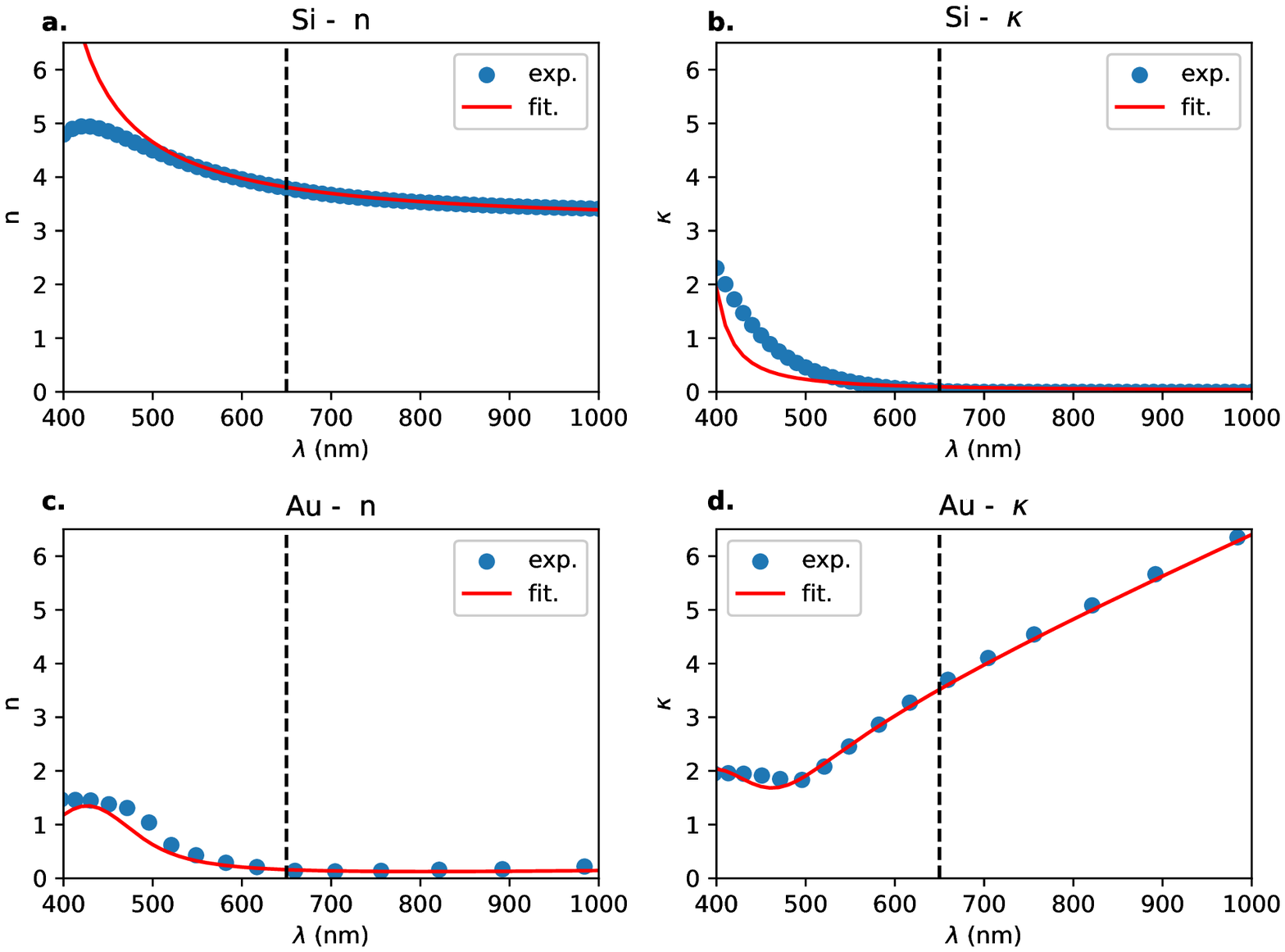}
    	\caption{Fitting of refractive index $n$ (\textbf{a}) and extinction coefficient $\kappa$ (\textbf{b}) of the amorphous silicon used for the fabrication of the nanoantennas, and of the gold used for the substrate (\textbf{c} and \textbf{d}). The fitting parameters for the analytical curves (red lines) are given in the Methods section in the main text. The experimental data (blue points) were obtained via ellipsometry measurements in the case of silicon, and taken from Ref.~\cite{johnson1972optical} in the case of gold. The wavelength of maximum emission of the quantum dots at $650\,$nm is also shown (vertical black dashed line).}
    	\label{asi}
    \end{figure*}

    We fit our experimental data of the optical constants of our amorphous Si (i.e. refractive index $n$ and extinction coefficient $\kappa$, obtained via ellipsometry measurements, shown in Figs.~\ref{asi}a, b), by a Lorentz model with $1$ pole (or oscillator), which is in a quantitative agreement with the experimental data for wavelengths $\lambda>550\,$nm. The complex permittivity $\varepsilon=\varepsilon'+\mathrm{i}\varepsilon''$, which is related to the optical constants by the relations $\varepsilon'=n^2-\kappa^2$ and $\varepsilon''=2n\kappa$, that we use in the case of Si reads:
    \begin{equation}
        \varepsilon_\text{Si}\left(\omega\right)=\varepsilon_{\infty , \text{Si}}\left[1-\frac{\omega^2_{p,\text{Si}}}{\omega^2-\omega^2_{0,\text{Si}}+\mathrm{i}\omega\gamma_{\text{Si}}}\right]
    \end{equation}
    with $\varepsilon_{\infty , \text{Si}}=1$, $\omega_{p,\text{Si}}=15\times 10^{15}\,$[rad/s], $\omega_{0,\text{Si}}=5\times 10^{15}\,$[rad/s], and $\gamma_{\text{Si}}=0.3\times 10^{15}\,$[rad/s].

    In the case of the Au substrate, we did not measure the optical constants and assumed that they correspond closely to the values measured by Johnson and Christy\cite{johnson1972optical}. In order to fit these experimental data, we used a Lorentz-Drude model with $2$ poles (oscillators), which is also in a quantitative agreement with the experimental data for wavelengths $\lambda>550\,$nm (see Figs.~\ref{asi}c, d). The complex permittivity that we use in the case of Au reads:
    \begin{equation}
        \varepsilon_\text{Au}\left(\omega\right)=\varepsilon_{\infty , \text{Au}}\left[1-\frac{\omega^2_{p,1,\text{Au}}}{\omega^2+\mathrm{i}\omega\gamma_{1,\text{Au}}}-\frac{\omega^2_{p,2,\text{Au}}}{\omega^2-\omega^2_{0,2,\text{Au}}+\mathrm{i}\omega\gamma_{2,\text{Au}}}\right]
    \end{equation}
    with $\varepsilon_{\infty , \text{Au}}=6$, $\omega_{p,1,\text{Au}}=5.37\times 10^{15}\,$[rad/s], $\gamma_{1,\text{Au}}=6.216\times 10^{13}\,$[rad/s], $\omega_{p,2,\text{Au}}=2.2636\times 10^{15}\,$[rad/s], $\omega_{0,2,\text{Au}}=4.572\times 10^{15}\,$[rad/s], and $\gamma_{2,\text{Au}}=1.332\times 10^{15}\,$[rad/s].

\clearpage

\bibliography{refs}